\documentclass[12pt,preprint]{aastex}

\def\LCDM{$\Lambda$CDM~}
\def\zp{z_{phot}}
\def\bzp{\bar{z}_{phot}}
\def\geqsim{\lower.73ex\hbox{$\sim$}\llap{\raise.4ex\hbox{$>$}}$\,$}
\def\leqsim{\lower.73ex\hbox{$\sim$}\llap{\raise.4ex\hbox{$<$}}$\,$}
\def\Mpch {~h^{-1}~{\rm Mpc}}

\newcommand{\myemail}{adm@astro.uiuc.edu}

\slugcomment{ApJ, in prep \today}
\shorttitle{Clustering Evolution of Photometrically-Classified QSOs}
\shortauthors{Myers et al.}

\begin{document}

\title{First Measurement of the Clustering Evolution of Photometrically-Classified Quasars}

\author{Adam D. Myers\altaffilmark{1,2}, Robert J. Brunner\altaffilmark{1,2}, Gordon T. Richards\altaffilmark{3,4}, Robert C. Nichol\altaffilmark{5}, Donald P. Schneider\altaffilmark{6}, Daniel E. Vanden Berk\altaffilmark{6}, Ryan Scranton\altaffilmark{7}, Alexander G. Gray\altaffilmark{8} and Jon Brinkmann\altaffilmark{9}}

\email{\myemail}

\altaffiltext{1}{Department of Astronomy,Ê 
University of Illinois at Urbana-Champaign,Ê 
Urbana, IL 61801}
\altaffiltext{2}{National Center for Supercomputing Applications,
Champaign, IL 61820}
\altaffiltext{3}{Princeton University Observatory, Princeton, NJ 08544}
\altaffiltext{4}{Dept. of Physics \& Astronomy, The Johns 
Hopkins University, 3400 N Charles St, Baltimore, MD 21218}
\altaffiltext{5}{ICG, Mercantile House, 
  Hampshire Terrace, University of Portsmouth, Portsmouth, P01 2EG, UK}
\altaffiltext{6}{Department of Astronomy and Astrophysics, 525 Davey Laboratory, Pennsylvania State University, University Park, PA 16802}
\altaffiltext{7}{Physics and Astronomy Department, University of Pittsburgh, 3941 O'Hara St., Pittsburgh, PA 15260}
\altaffiltext{8}{Robotics Institute, Carnegie Mellon, 3128 Newell-Simon Hall, 5000 Forbes Ave, Pittsburgh, PA 15213}
\altaffiltext{9}{Apache Point Observatory, P.O. Box 59, Sunspot, NM 88349}

\begin{abstract}
   
  We present new measurements of  the quasar angular autocorrelation function
  from a sample of $\sim$80,000 photometrically-classified
  quasars taken from the First Data Release of the Sloan Digital Sky
  Survey. We find a best-fit model of $\omega(\theta) =
  (0.066\pm^{0.026}_{0.024})\theta^{-(0.98\pm0.15)}$ for the angular correlation
  function, consistent with estimates of the slope from spectroscopic quasar surveys. We show that only models with little or no evolution in the clustering of
  quasars in comoving coordinates since a median redshift of
  $z\sim1.4$ can recover a scale-length consistent with local galaxies and Active Galactic Nuclei (AGNs).  A model with little evolution of quasar clustering in comoving coordinates is best explained in the current cosmological paradigm by rapid evolution in quasar bias. We show that quasar biasing must have changed from $b_Q\sim3$ at a (photometric) redshift of $\bzp=2.2$ to $b_Q\sim1.2-1.3$ by $\bzp=0.75$. Such a rapid increase with redshift in biasing implies that quasars at $z\sim2$ cannot be the progenitors of modern $L^*$ objects, rather they must now reside in dense environments, such as clusters.  Similarly, the duration of the UVX quasar phase must be short enough to explain why local UVX quasars reside in essentially unbiased structures.  Our estimates of $b_Q$ are in good agreement with recent spectroscopic results \citep{Cro05}, which demonstrate that the implied evolution in $b_Q$ is consistent with quasars inhabiting halos of similar mass at every redshift.
Treating quasar clustering as a bivariate function of both redshift and luminosity, we find no evidence for luminosity dependence in quasar clustering, and that redshift evolution thus affects quasar clustering more than changes in quasars' luminosity.  Our results are robust against a range of systematic uncertainties. We provide a new method for quantifying stellar contamination in photometrically-classified quasar catalogs via the correlation function.

\end{abstract}


\section{Introduction}

Determining the distribution of matter (baryonic and dark),
as a function of redshift, is a fundamental goal of cosmology, providing important information on the content of the Universe.
Unfortunately, the majority of this matter is non-baryonic and we are forced to use tracers, like galaxies and quasars, to infer its presence.  Study of these populations introduces the added complication of determining how such tracers
are biased compared with underlying dark matter.  Nevertheless, vast resources are devoted to mapping the distribution of such tracers to help infer the distribution of matter
in the Universe.

In the local Universe ($z<0.1$), the distribution of galaxies (and
thus inferred matter) has become increasingly constrained due to large
redshift surveys such as the 2dF Galaxy Redshift Survey \citet{Col01} and Sloan Digital Sky Survey (SDSS; \citealt{Gun98,Lup99,Yor00,Hog01,Smi02,Sto02,Pie03,Ive04}). Such surveys are now able to measure the density of dark matter to better than 10\% and detect quite subtle features in the galaxy distribution, such as ``baryon acoustic oscillations'' (see, e.g., \citealt{Teg04, Eis05, Col05}). In the distant Universe ($z\simeq1$), the distribution of galaxies is less
constrained, as it requires extensive investments of telescope time on
8-meter class telescopes (e.g., DEEP2, \citealt{Dav03}). At even higher
redshifts ($z>1.5$), our knowledge of galaxies is limited
by the redshifting of the bulk of their luminosity to infrared \& submillimeter
wavelengths (e.g., the GDDS survey, \citealt{Abr04}).

When studying the distribution of matter at high redshifts
($z>1$), quasi-stellar objects (quasars or QSOs) are a better
tracer than galaxies, as they are extremely luminous and can
be identified from current multi-color optical imaging out to a redshift
of $z\simeq6.5$ \citep{Fan03}. However, there are problems with using quasars as tracers of dark matter, for instance: 1) It has long been unclear how quasars are physically related to the underlying dark matter halos they inhabit, thus leading to uncertainty about their
biasing schema. 2) Existing photometrically-selected quasar samples suffer from significant
stellar contamination ($\sim50\%$) and thus require
laborious follow-up spectroscopy (see, e.g., \citealt{Cro04}).

In recent years, these two issues have begun to be addressed. In the case of
quasar bias, it has become increasingly clear that all massive
galaxies possess a central supermassive black hole \citep{Ric98}, which has a mass correlated to that of its parent dark matter halo (see \citealt{Fer00,Geb00} etc.).
Under the hypothesis that the quasar phenomenon is driven by these
supermassive black holes, it is likely that quasar properties are
linked to the evolution of underlying dark matter, as demonstrated in high-resolution simulations (e.g., \citealt{DiM04}).  There have also been recent advances
in the algorithms used to identify quasars. For example, \citet{Ric04} have developed a
Bayesian method to separate quasars from stars in the
4-dimensional color-space of the SDSS, with nominal stellar contamination as low as
5\% for redshifts of 0.2~\leqsim~z~\leqsim 2.4.

The distribution of objects in the Universe is commonly quantified
using the two-point correlation function \citep{Tot69}, or its Fourier counterpart,
the power spectrum of density fluctuations. The 2dF QSO Redshift Survey (henceforth 2QZ; \citealt{Cro04}) has provided the most precise estimates (to date) of both the quasar power spectrum \citep{Out03} and two-point correlation function \citep{Cro05}. In particular, \cite{Cro05} find the biasing of quasars evolves from $b_Q\sim4.4$ at $z=2.48$ to
$b_Q\sim1.1$ at $z=0.53$ (see also \citealt{Por04}), consistent with observations that local AGNs are unbiased with respect to the normal galaxy population (e.g., \citealt{Wak04}).  Together, these results constrain the mass and evolution of the dark matter halos that harbor QSOs, as well as the duration of the quasar phase (e.g., \citealt{Gra04,Hop05}).

In summary, it is now possible to employ photometrically-classified quasars to investigate quasar bias.  We introduce just such an application in this paper, in which we investigate the evolution of quasar clustering using the largest published sample of
photometrically-classified quasars \citep{Ric04}. We use photometric
redshift estimates \citep{Wei04}, inverting the angular
correlation function to estimate the amplitude of quasar clustering in real-space and the evolution of quasar bias with redshift. After discussing our data, and possible systematic uncertainties, we will measure the angular autocorrelation of our quasar sample.  We will then present the first estimate of the clustering evolution of photometrically-classified quasars, also exploring luminosity-dependent clustering. Unless otherwise specified, cosmological modeling in this paper assumes $(\Omega_m, \Omega_\Lambda,h\equiv H_0/100{\rm km~s^{-1}~Mpc^{-1}}) = (0.3,0.7,0.7)$; consistent with WMAP results \citep{Spe03}, and magnitudes are corrected for Galactic extinction (using the dust maps of \citealt{Sch98}).

\section{Data and Methodology}

\subsection{The KDE Data}

The data analyzed in this paper are from the photometrically-classified sample of \citet{Ric04}, which we henceforth call the KDE (Kernel Density Estimation) catalog.  The KDE catalog is drawn from point sources with $u-g < 1$, (observed) $g \geq 14.5$ and (dereddened) $g < 21$, that appear in the SDSS First Data Release (DR1; \citealt{Aba03}). Separations, in 4-dimensional color-space, from the (spectroscopically-confirmed) quasar and stellar loci, are determined for each source.  A Bayesian technique is used to classify each object as either ``QSO'' or ``star''.  The resulting set of 100,563 quasar candidates is thus UVX-and-magnitude-limited in such a way as to be broadly comparable to the 2QZ.  In general, we will refer to objects from the KDE catalog as QSOs, even though the vast majority of the KDE data have not been spectroscopically confirmed as quasars. The KDE catalog includes a photometric redshift (photoz) for each QSO (see \citealt{Wei04}).

\subsection{The Random Catalog}
\label{sec:rancat}

Estimating angular correlation functions requires a sample of random points that have the same angular selection function as the data surveyed (neglecting correlations that are due to QSOs tracing cosmological structure).  To construct a random catalog that mimics the KDE data, we create a large set of random points distributed over the DR1 area. Points that fall in any SDSS imaging mask\footnote{http://www.sdss.org/dr1/products/image/use\_masks.html} are discarded. We assign each random point the seeing value of the nearest PhotoPrimary object in the SDSS database\footnote{http://cas.sdss.org} and its absorption value from the Galactic dust maps of \citet{Sch98}. Broad sky coverage of the KDE data is shown in Figure~4 of \citet{Ric04}, although the sample we use is also cut to the DR1 theoretical footprint, which discards 3.4\% of the KDE data. The SDSS theoretical footprint differs from the actual sky coverage in the South, as curvature of the coordinate system forces drift-scanning beyond the targeted stripes.

\subsection{Correlation Function and Error Estimation}

We construct the two-point angular correlation function ($\omega$) from counts of data-data, random-data and random-random pairs, via the estimator of \citet{Lan93}. We use logarithmic bins centered at each angular separation ($\theta$).  The estimator is

\begin{equation}
\omega(\theta) = \frac{QQ(\theta) - 2QR(\theta)}{RR(\theta)} + 1
\label{eqn:LScorr}
\end{equation}

\noindent where $Q$ denotes a data point and $R$ denotes a random point (see section~\ref{sec:rancat}).  We use a random catalog 100 times larger than the data catalog, normalizing the pair counts accordingly, and only quote results for bins that contain at least 10 data points.

We estimate errors using jackknife resampling \citep{Scr02}, which performs well across a range of scales \citep{Zeh02,Mye05}.  The jackknife method is to divide the data into $N$ pixels, then create $N$ subsamples by neglecting each pixel in turn.  If we denote subsamples by the subscript $L$ and recalculate $\omega_L$ via Equation~\ref{eqn:LScorr}, then the jackknife error, $\sigma_{\omega}$ is

\begin{equation}
\sigma_{\omega}^2(\theta) =
\sum_{L=1}^{N}\frac{RR_{L}(\theta)}{RR(\theta)}\left[\omega_{L}(\theta)
- \omega(\theta)\right]^2
\label{eqn:JACKerr}
\end{equation}

\noindent  The $RR_L/RR$ term \citep{Mye05} weights by the different numbers of objects expected, due to holes, poor seeing or pixels that extend beyond the DR1 boundary.  Throughout this paper, we jackknife-resample using 1 deg$^{2}$ pixels, sampling across thousands of realizations. 

\begin{figure}
\plotone{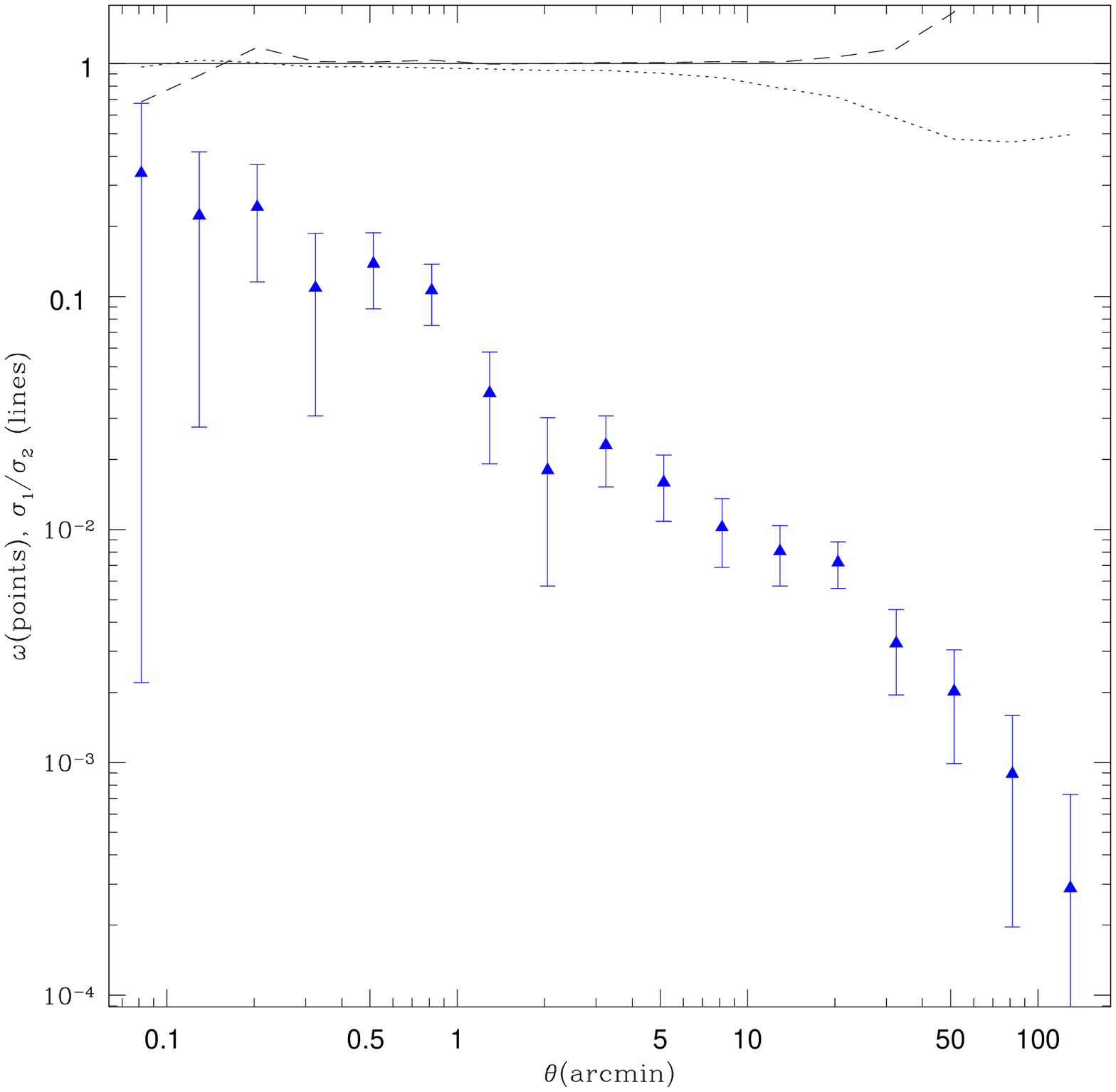}
\caption{The autocorrelation of all data from the KDE catalog with error comparison.  The points plot the autocorrelation estimated using Equation~\ref{eqn:LScorr}, which we have independently verified are in excellent agreement with what is measured using the approach of \cite{Scr02}.  We plot jackknife errors but have also computed Poisson and pixel-to-pixel errors.  The lines plot the ratio of Poisson to jackknife errors (dotted), and the ratio of pixel-to-pixel to jackknife errors (dashed).\label{fig:allkde}}
\end{figure}

In Figure~\ref{fig:allkde}, we display the autocorrelation of all objects in the KDE catalog, with jackknife errors. We also display the jacknife errors in ratio to: 1) Poisson errors (calculated via $\sigma_{\omega}^2 = 2(1+\omega)^2/QQ$ as only half of the $QQ$ pairs are independent) and; 2) pixel-to-pixel errors (see \citealt{Mye03} for more on these errors). The error ratios in Figure~\ref{fig:allkde} illustrate that: 1) Poisson errors become systematically smaller than jackknife errors as the scale increases, and that; 2) the pixel-to-pixel error becomes ill-defined on the scale of a pixel (1 deg in Figure~\ref{fig:allkde}). We therefore use the jackknife error throughout this paper.

\subsection{Modeling}

To fit models to the angular correlation function, we use a power-law \citep{Pee80}

\begin{equation}
\omega(\theta) = A\theta^{-\delta}
\label{eqn:projmod}
\end{equation}

\noindent of amplitude $A$. The slope, $\delta$, is canonically found to be $0.7-0.8$ for galaxies (e.g., \citealt{Con02}) although the amplitude depends on the bias of the galaxy type.  \citet{Cro05} suggest that $\delta$, when averaged over $1-100\Mpch$ scales, ranges from  $0.65-0.85$ for QSOs ($\Lambda$-dominated cosmology) depending on whether redshift-space distortions are included in the model; they further note that a single power-law may not fairly represent QSO clustering, since QSO correlation estimates are capable of probing the non-linear, intermediate and linear regimes, which, according to perturbation theory, have different clustering amplitudes.

\section{Sources of Systematic Error}

We now address possible contaminants of our clustering analysis.  Given the large areas we sample, we neglect the effect of the integral constraint, as it will be an order of magnitude or more smaller than our typical error (c.f. \citealt{Scr02,Con02}).

\subsection{Star/Galaxy Classification Errors}

\begin{figure}
\plotone{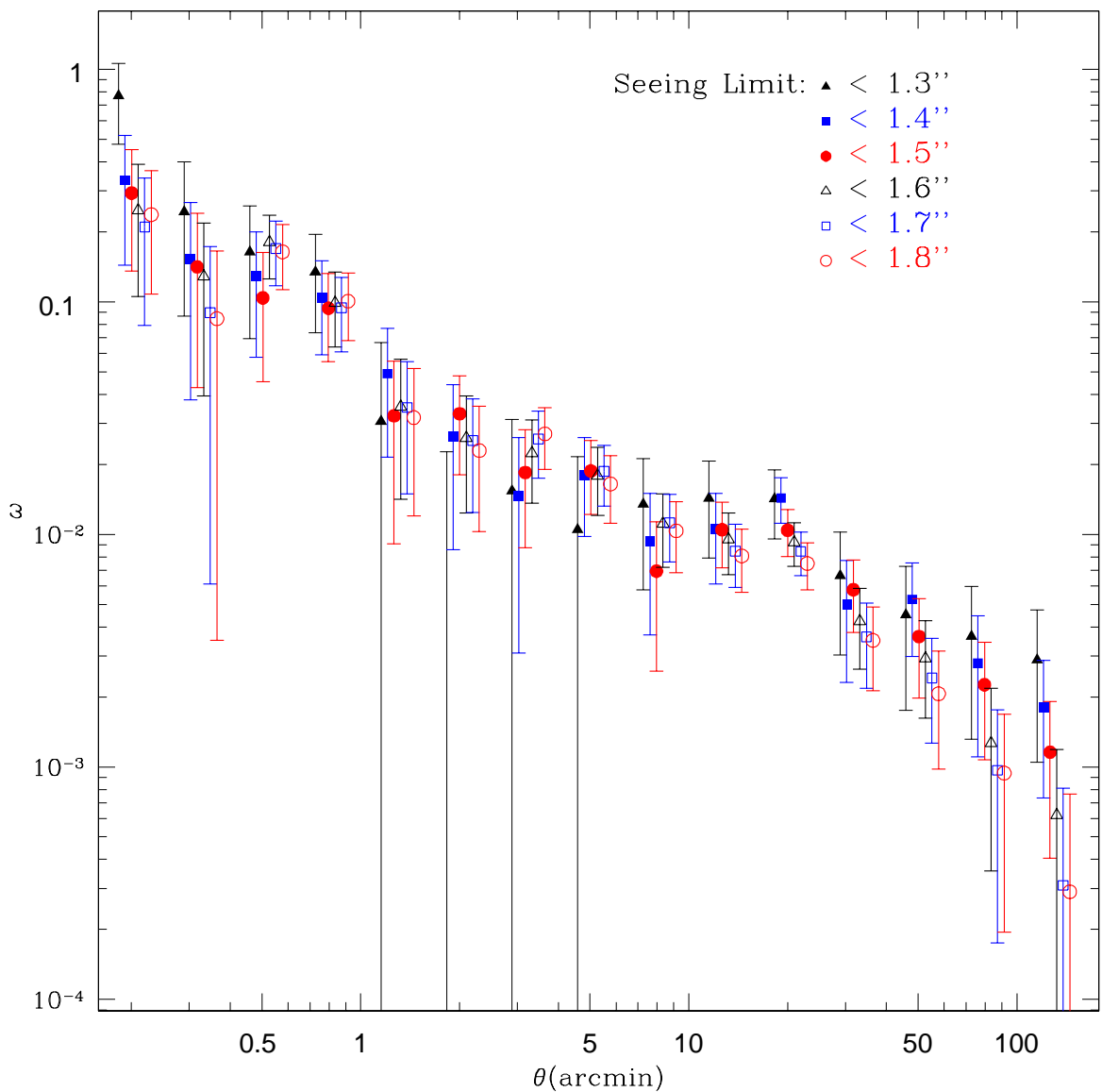}
\caption{The effect of seeing on the KDE QSO autocorrelation. The autocorrelation of all KDE objects is plotted for a range of seeing cuts. Seeing in this plot is measured in the $g$ band.  Note that the most liberal cut of 1.8~arcsec or better is effectively the same as making no seeing cut.  All errors are jackknifed. Points have been offset within each bin to aid clarity. \label{fig:seecomb}}
\end{figure}

If galaxies that are normally resolved are imaged in poor seeing, they may be misclassified as point sources and included in the KDE catalog, impacting autocorrelation measurements.  In Figure~\ref{fig:seecomb} we display the autocorrelation as a function of ($g$-band) seeing for the entire KDE catalog. Seeing cuts of $< 1.8$ and $< 1.3$~arcsec remove, respectively, $\sim$5\% and $\sim$70\% of the KDE data.  KDE objects with 2QZ spectroscopic matches suggest that contamination by galaxies not resolved in $1.3$~arcsec seeing is extremely small ($\leqsim0.2\%$). We have used the $\chi^2$ statistic to estimate the amplitude and slope of the correlation function for the data shown in Figure~\ref{fig:seecomb}.  Though there are very weak trends with seeing on the largest scales we consider, they are consistent with no dependence given the errors.  As the hypothesis that seeing does not influence our model of QSO clustering is allowed within the errors, and intrinsic fluctuations in our correlation estimates are far larger, we enforce no seeing constraints throughout this paper.  


\subsection{Galactic Extinction}

\begin{figure}
\plotone{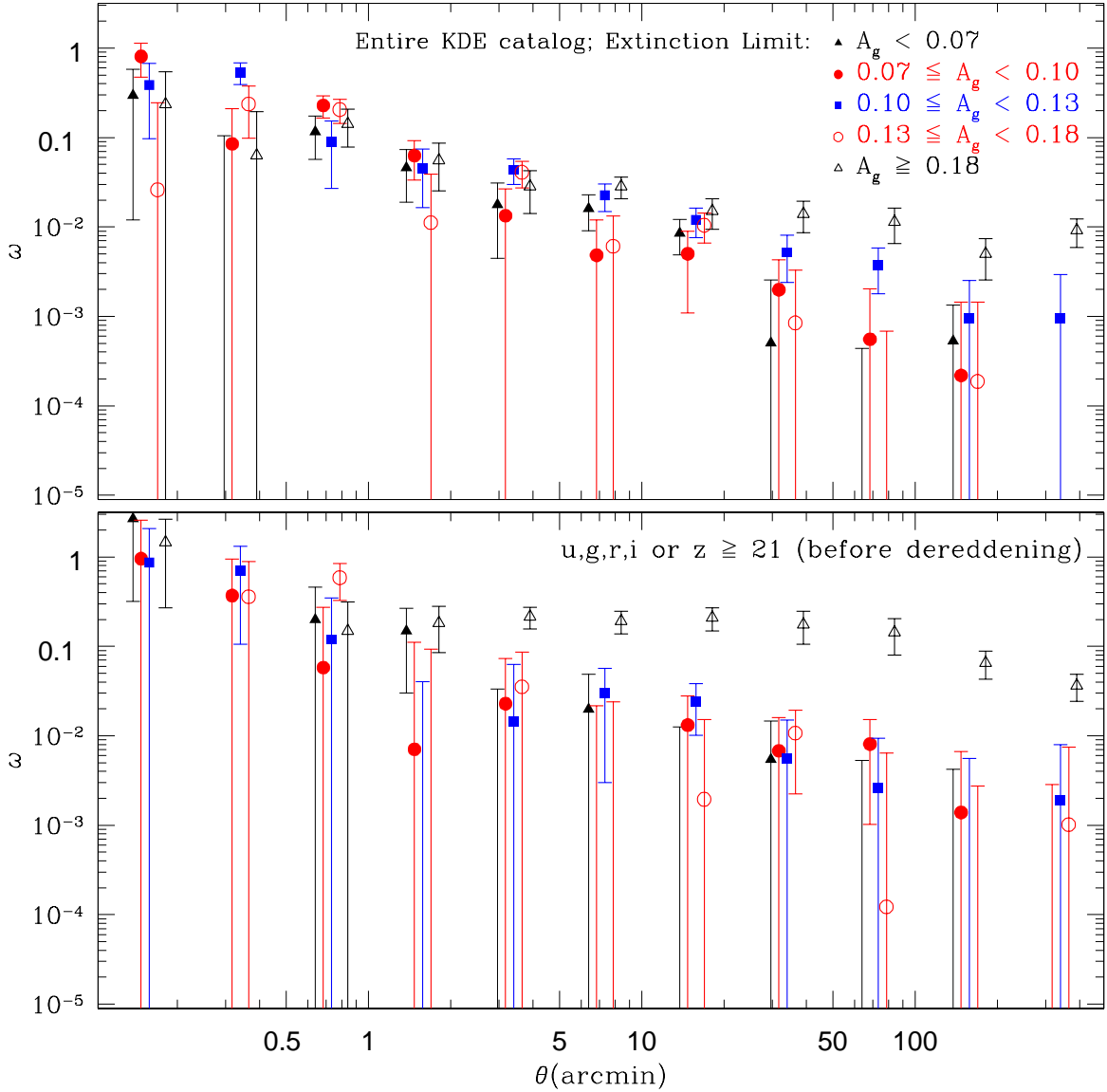}
\caption{The effect of absorption by Galactic dust on the KDE autocorrelation.  The upper panel shows the KDE autocorrelation binned by galactic absorption in $g$, with $\sim$20,000 objects in each bin.  Clearly absorption of $A_g \geq 0.18$ introduces spurious power on large scales.  The lower panel repeats the analysis, in the same bins of $A_g$, for objects that are both faint and highly reddened, by considering objects observed fainter than 21 in any SDSS band (i.e. with no correction to magnitude for Galactic dust).  There are $\sim$5,000 objects per bin.  Faint, highly-obscured objects introduce large-scale clustering and are the main culprit in causing the effect for $A_g \geq 0.18$.  Objects observed with magnitudes $> 21$ that are not heavily obscured by Galactic dust display no such effect. All errors are jackknifed. Points have been offset within each bin to aid clarity. \label{fig:dust}}
\end{figure}

Dust in our Galaxy causes QSO numbers to fluctuate but the dereddened magnitude limit of $g < 21$ imposed on the KDE catalog, coupled with the SDSS (95\% detection) limit of $g < 22.2$ means that few QSOs should be obscured from the KDE sample.  In the upper panel of Figure~\ref{fig:dust} we plot the autocorrelation of KDE objects as a function of absorption ($A_g$). The autocorrelation in $A_g$ bins resembles that of the entire catalog except for $A_g \geq 0.18$, where Galactic extinction introduces excess clustering on large scales.  Our binning in $A_g$ is chosen to ensure sufficient objects ($\sim$20,000) in each bin to study the clustering signal with some significance. We note that our $A_g < 0.18$ cut is stricter than the $A_r < 0.2$ cut suggested by \citet{Scr02}, so it is possible that a less strict cut in $A_g$ exists that still rejects most of the dust-induced excess clustering.  However, changing the bin resolution does not affect our conclusion that a cut of $A_g < 0.18$ is sufficient to remove all of the large-scale power, and a less strict cut merely risks introducing a systematic without significantly improving the statistical precision of any clustering measurements. The lower panel of Figure~\ref{fig:dust} illustrates that the main cause of the dust-induced excess clustering power is objects that are both observed to have faint magnitude and that suffer high Galactic absorption.  It is unclear if these objects exhibit large-scale clustering due to correlations between dust enshrouded regions of the Galaxy or because they are misclassified stars.  Throughout our analyses, the KDE catalog and the random catalog are cut to regions with $A_g < 0.18$, discarding $\sim$20\% of the sample.

\subsection{Stellar Contamination}
\label{sec:stelcon}

Based on classifying simulated QSOs, \citet{Ric04} find that the KDE technique is 95\% efficient, although their Figure~6 suggests this efficiency is magnitude-dependent.  Once other sources of potential clustering contamination, such as seeing and dust, have been eliminated, we can use clustering measurements to independently test this 95\% efficiency claim.  Further, we can test how 5\% stellar contamination would impact the quasar autocorrelation.

Under the substitution $Q \to aQ+(1-a)S$, where $a$ is the efficiency, or fraction of
correctly classified quasars, it can be shown that
Equation~\ref{eqn:LScorr} becomes
\begin{equation}
\omega(\theta) = a^2\omega_{QQ}(\theta) + (1-a)^2\omega_{SS}(\theta) + \epsilon(\theta)
\label{eqn:stelcon}
\end{equation}
where $\omega_{QQ}$ and $\omega_{SS}$ are the intrinsic
autocorrelation of QSOs and stars, respectively, and $\epsilon$ is a
tiny offset arising from cross-terms
\begin{equation}
\epsilon=2(a^2-a)\left[\frac{QR+SR-QS}{RR}-1\right]
\end{equation}
As we model the uncontaminated quasar and star distributions,
$QR/RR$, $QS/RR$ and $SR/RR$ should be $\sim 1$, so $\epsilon
\sim 0$. We estimate limits on these cross-terms using: 1) (for quasars) all ($A_g < 0.18$) KDE objects and; 2) DR1 stellar sources with $16.9 \leq g < 17.1$, a range chosen to both match the quasar sample size  and to ensure that few quasars are present to contaminate the stellar clustering signal.  We have checked if $\omega_{SS}$ is constant across the range $17 \leq g < 21$, finding a slightly smaller amplitude at $g\sim21$, which is insufficient to change any of our conclusions throughout this section, and may, in any case, be due to quasar contaminants. Over 1-60~arcmin scales we find $|1-(QR/RR)|<0.005$, $|1-(SR/RR)|<0.008$ and $|1-(QS/RR)|<0.03$. Summing these limits in quadrature implies $|\epsilon| < 0.063(a^2-a)$. As $0\leq a\leq1$, we conclude $|\epsilon| < 0.016$ as a strong upper limit. In this section, we assume $\epsilon=0$ but $\epsilon=0.063(a^2-a)$ would not affect our conclusions.

\begin{figure}
\plotone{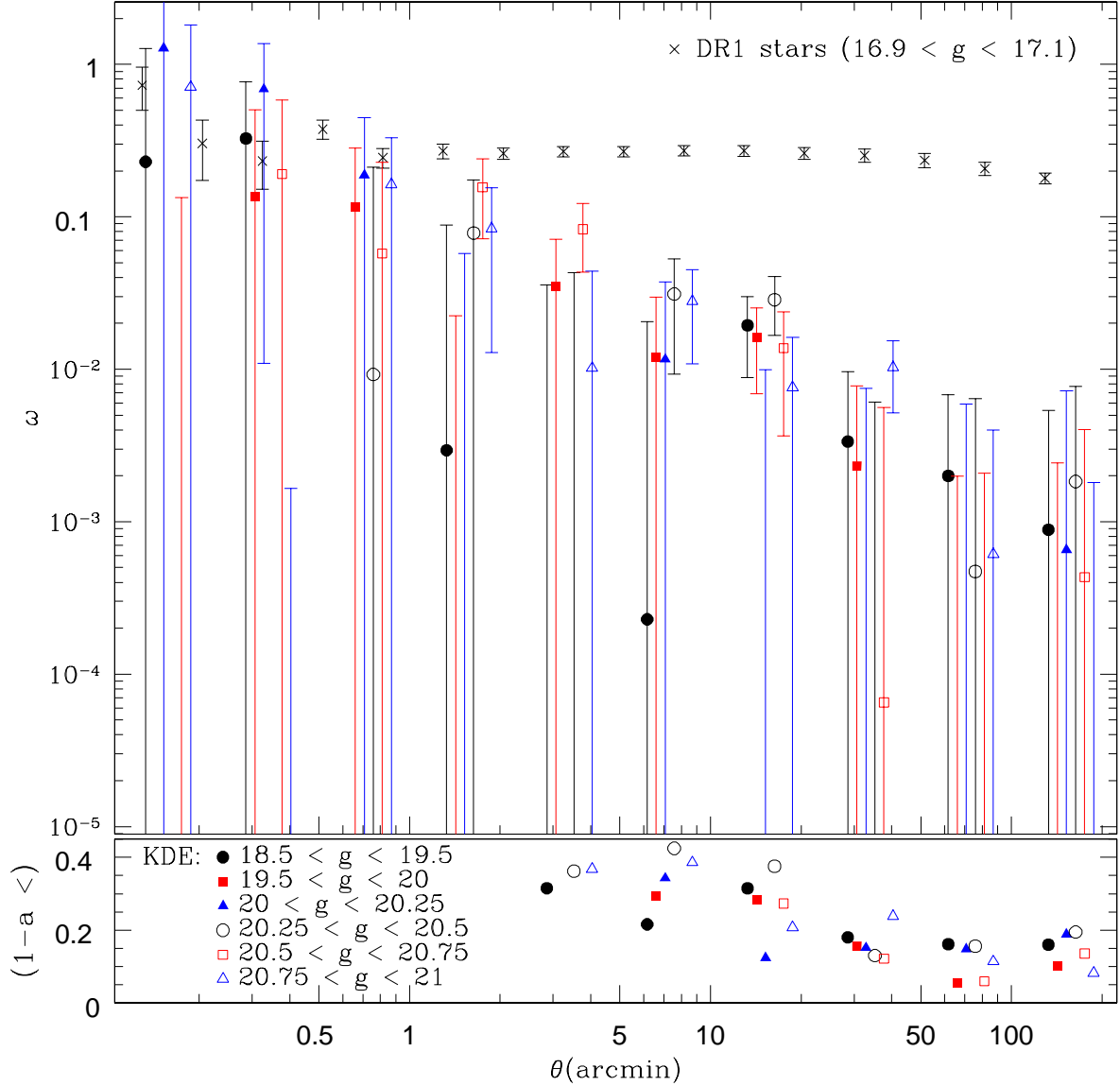}
\caption{To search for magnitude-dependent stellar contamination in the KDE catalog we consider, in the upper panel, the autocorrelation of ($A_g < 0.18$) KDE QSOs as a function of $g$ magnitude.  The KDE data have been divided into 6 bins of approximately equal numbers, containing $\sim$12,600, $\sim$14,100, $\sim$9,700, $\sim$11,400, $\sim$13,300 and $\sim$15,000 QSOs from brightest to faintest, respectively. In the lower panel, the plotted points represent {\it $1\sigma$ upper limits} on the stellar contamination ($1-a$ in Equation~\ref{eqn:stelcon}).  These limits are derived using $\omega_{SS}$ estimated from star-like objects in DR1 that have magnitude in the range $16.9 \leq g < 17.1$ (plotted as crosses in the upper panel), and taking $\omega_{QQ}$ from all ($A_g < 0.18$) KDE objects. Points have been offset within each bin to aid clarity. All errors are jackknifed.
\label{fig:stelcon}}
\end{figure}

In the upper panel of Figure~\ref{fig:stelcon} we show the autocorrelation of ($A_g < 0.18$) KDE objects in bins of $g$ magnitude.  There is no evidence of magnitude-dependent effects.  Stellar contamination would be signified by similar clustering on all scales, since a degree probes tens of parsecs for typical Galactic stars detected in the SDSS, but tens of \textit{Mega}parsecs at QSO distances.  For comparison we also plot the autocorrelation for stellar sources in DR1 in the range
$16.9 \leq g < 17.1$. At 30~arcmin $\omega_{SS}\sim0.25$, dropping to $\sim0.18$ at 2~degrees.  We can use Equation~\ref{eqn:stelcon} to quantify stellar contamination ($1-a$) in the KDE sample.  In the lower panel of Figure~\ref{fig:stelcon} we plot the {\it $1\sigma$~upper limits} on $1-a$, taking $\omega_{QQ}$ from all ($A_g < 0.18$) KDE objects.  Though any scale is valid for measuring the upper limit on $1-a$, the effects of any stellar contamination should be most apparent on larger scales (where $\omega_{QQ} \rightarrow 0$). The lowest upper limit measured in each bin of $g$ is 15.5\%, 5.6\%, 13.0\%, 6.0\%, 8.2\%, 12.3\% for bins with $\bar{g}$ of 19.10, 19.76, 20.13, 20.38, 20.63, 20.88, respectively, and the lower limit is consistent with zero in every bin.

As a further test, we construct a sample of spectroscopic matches to the 2QZ. Unlike \citet{Ric04}, we do not consider matches to the DR1 quasar catalog (\citealt{Sch03}; henceforth DR1QSO), as it may be biased to quote efficiencies from data that the KDE algorithm was trained on.  We also ignore matches to SDSS DR2 (\citealt{Aba04}), which are selected similarly to DR1QSO.  Our matches consist of quality ``1'' 2QZ QSOs (see \citealt{Cro04}) within 2~arcsec of a KDE object.  For ($A_g < 0.18$) KDE objects with 2QZ matches, 
we find efficiencies of 97.9\%, 97.3\%, 95.7\%, 93.9\%, 90.7\%, 88.8\%, for $\bar{g}$ as above, and 95.1\% across the full magnitude range. This is consistent with both \citet{Ric04} and our measurements of $a$ from clustering.  We note that objects targeted by several independent UVX methods are more likely to be quasars (the ratio of quasars to non-quasars for quality ``1'' 2QZ objects that overlap DR1 is 57:43). 

If the efficiency of the KDE sample is as high as our analyses suggest, we can measure $\omega_{QQ}$ without stellar contamination having an impact. Taking $\omega_{SS}(2$~degrees)=0.18, a contamination of 5\% would cause an equal contribution from quasars and stars for $\omega_{QQ}\sim0.0005$, which, will be comparable with our error on $\omega$ at 2~degrees.  Similarly, a contamination of 10\% would cause an equal contribution for $\omega_{QQ}\sim0.0022$, which is negligible, given that when we bin our sample by magnitude the smallest error on $\omega(2$~degrees) is 0.0074.

\section{The Projected Clustering of QSOs}

\subsection{The KDE QSO Autocorrelation}
\label{sec:fullcorr}

\begin{figure}
\plotone{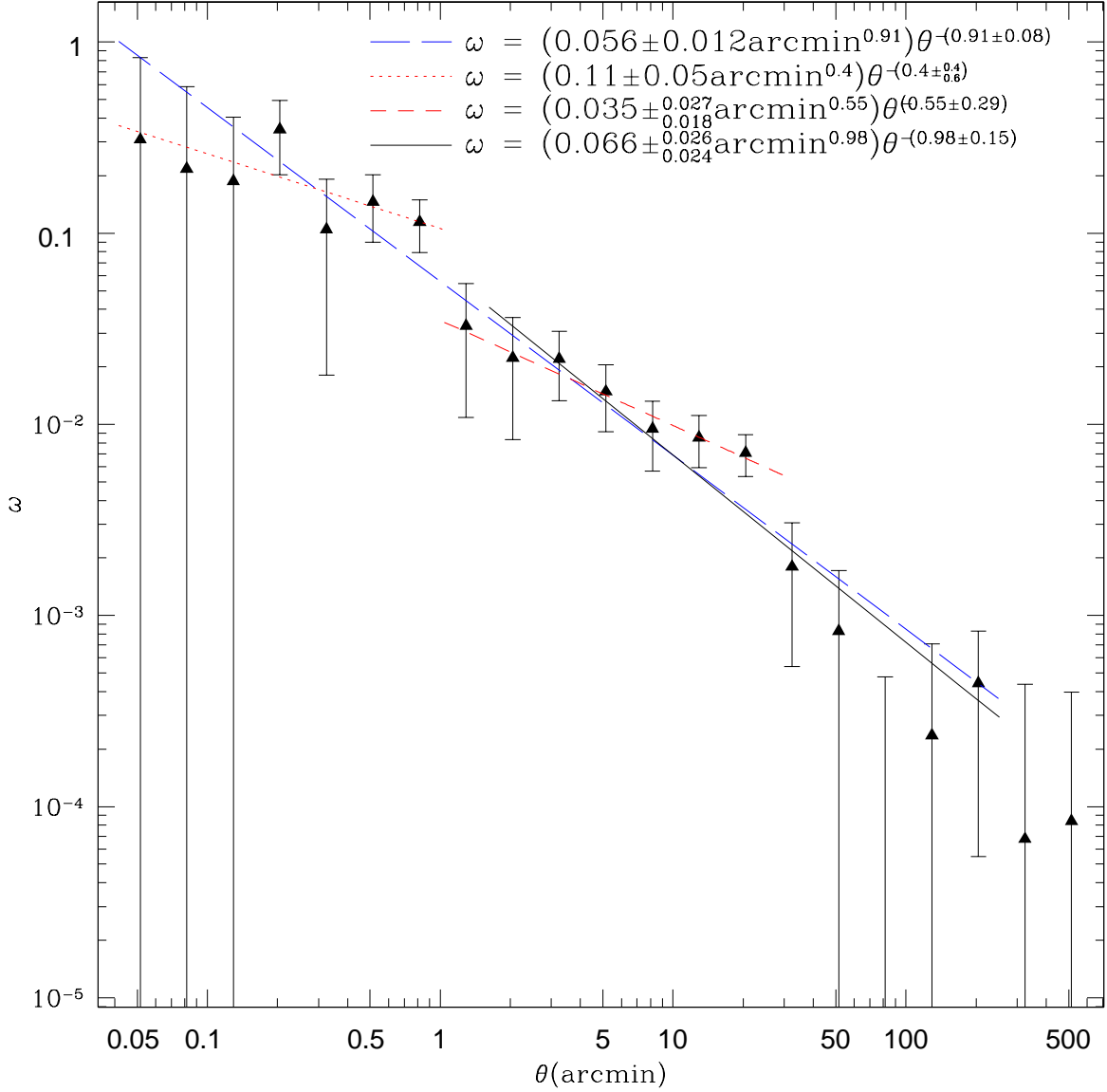}
\caption{The autocorrelation of $A_g < 0.18$ KDE objects.  Points are plotted with jackknife errors, and binned logarithmically. The best-fitting power-law model across ``all'' scales ($0.04<\theta<250$~arcmin) is displayed as a long-dashed line. There is marginal evidence for a break in the power-law at $\sim$25~arcmin.  The lines display the best-fitting power-law model for the scales over which they are plotted; $0.04<\theta<1$~arcmin (dotted line), $1<\theta<25$~arcmin (short-dashed line) and $2<\theta<250$~arcmin (solid line). A scale of 10~arcmin is $\sim$3.5$\Mpch$ at the median redshift of the ($A_g < 0.18$) KDE sample.
\label{fig:extended}}
\end{figure}

In Figure~\ref{fig:extended} we display the autocorrelation of our ($A_g < 0.18$) KDE sample, together with best-fitting models (from Equation~\ref{eqn:projmod}).  The long-dashed line, a fit across ``all'' scales, is marginally rejected, with $P(<\chi^2)=0.27$, which is expected as a single power-law is not a good fit to either spectroscopic quasar samples or CDM models (see, e.g., \citealt{Cro05}).  To the eye, Figure~\ref{fig:extended} suggests breaks at $\sim$1~arcmin and $\sim$25~arcmin.  A small-scale break is worth investigating, as \citet{Zeh04} find power-law departures on $1-2\Mpch$ scales for the projected galaxy autocorrelation.  However, based on Poisson statistics, which are valid on small scales, (see Figure~\ref{fig:allkde}), there is no evidence for a break at 1~arcmin, as the data suggest $\omega(<1~{\rm arcmin})=0.131\pm0.036$ and integrating the fitted model yields 0.103.  If we model with a break at 1~arcmin, the fits are only slightly improved, with $P(<\chi^2)=0.80$ for $\theta < 1$~arcmin (the dotted line in Figure~\ref{fig:extended}) and $P(<\chi^2)=0.36$ for $\theta > 1$~arcmin (similar to the solid line in Figure~\ref{fig:extended}).  There {\it is}, however, marginal evidence for a break at 25~arcmin, as $\omega(<25~{\rm arcmin})=0.0086\pm0.0013$, compared to 0.0055 for the model.  Corrected for the ratio between Poisson and jackknife errors (see~Figure~\ref{fig:allkde}), this is a $1.7\sigma$ fluctuation.  Further, a power-law fit over 1-25~arcmin (the short-dashed line in Figure~\ref{fig:extended}) provides an excellent fit of $P(<\chi^2)=0.98$. We intend to repeat our analyses with a larger sample of photometrically-classified quasars, drawn from SDSS DR4, and will soon know whether this break persists in a larger sample.

The regime that we will consider is $\theta~\geqsim~2$~arcmin ($\geqsim~0.75\Mpch$ at the sample's median redshift, $z\sim1.4$, as calculated from KDE objects with spectra in DR1QSO). We study these scales as, on average, they are dominated by points at $\theta~\geqsim~30$~arcmin ($\geqsim~10\Mpch$; which should be in the linear regime of clustering) but can still provide meaningful constraints from the KDE sample. Fits to the data are statistically unchanged by fitting from 2~arcmin out to any maximum scale in (at least) the range 40-250~arcmin ($\sim$14-89$\Mpch$), as we will indirectly demonstrate in section~\ref{sec:gausswide}. Over scales of $2 < \theta < 250$~arcmin ($\sim$0.75-89$\Mpch$), our best-fitting power-law model has a slope of $\delta = 0.98 \pm 0.15$ consistent with \citet{Cro05}, who find (accounting for distortions that affect clustering along the redshift coordinate), a nearly acceptable power-law fit with slope $\gamma=\delta+1=1.866\pm0.060$ over $1-100\Mpch$.  Finally, we note that the largest-scale points we plot in Figure~\ref{fig:extended} are statistically consistent with being anti-correlated, in agreement with CDM models, which go negative around $70-100\Mpch$ (e.g., \citealt{Cro05}).  Further, the bin plotted at $\sim$200~arcmin spans the range $56-140\Mpch$ and is tantalizingly higher than adjacent bins.  This scale is consistent with that expected for a baryon peak (see, e.g., \citealt{Eis05}) but the higher amplitude of the bin at $\sim$200~arcmin is certainly not statistically significant in the current sample.


\subsection{Limber's Equation and the Evolution of the Quasar Correlation Function}
\label{sec:Limb}

When the angular correlation is expressed as in Equation~\ref{eqn:projmod}, the de-projected spatial correlation function can be written \citep{Pee80}

\begin{equation}
\xi(r,z) = \left(\frac{r}{r_0(z)}\right)^{-\gamma}=\left(\frac{r}{r_0}\right)^{-\gamma}(1+z)^{-(3+\epsilon)}
\label{eqn:realcorr}
\end{equation}

\noindent where $\gamma$ is the power-law slope, $r_0$ is the local spatial scale-length, and $\epsilon$ parameterizes clustering evolution.  In general, for $\epsilon < 0$, clustering diminishes with cosmic time, meaning objects with high redshift were more clustered.  


The spatial correlation function can be integrated to yield its angular projection \citep{Lim53}.  In the small angle approximation ($\theta \ll 1~{\rm radian}$), unknowns in Equations~\ref{eqn:projmod} and \ref{eqn:realcorr} can be related (see \citealt{Pee80} for a full derivation)

\begin{eqnarray}
\delta&=& \gamma -1\\
A &=&  H_\gamma \frac{\int^{\infty}_0({\rm d}N/{\rm d}z)^2E_z(1+z)^{\gamma-(3+\epsilon)}\chi^{1-\gamma}{\rm d}z}{\left[\int^{\infty}_0({\rm d}N/{\rm d}z){\rm d}z\right]^2} r_{0}^\gamma
\label{eqn:deproj}
\end{eqnarray}

\noindent where $H_\gamma=\Gamma(0.5)\Gamma\left(0.5[\gamma-1]\right)/\Gamma(0.5\gamma)$, $\Gamma$ is the gamma function, $\chi$ is the radial comoving distance, {\rm d}N/{\rm d}z is the redshift selection function, and $E_z = H_z/c = {\rm d}z/{\rm d}\chi$. Strictly, $\chi$ should be the angular, or transverse, comoving distance, however, in our chosen, flat cosmology, radial and transverse comoving distances are equivalent. The Hubble Parameter can be found via 

\begin{equation}
H_z^2 = H_0^2\left[\Omega_m(1+z)^3+\Omega_\Lambda\right]
\end{equation}

Perturbation theory suggests the mass correlation function is scale-dependent, transitioning from linear to non-linear scales at $10-20\Mpch$ at $z\sim0$ (e.g., \citealt{Ham02}), and at smaller scales at higher redshift.  In the highly non-linear regime ($\ll10\Mpch$), the correlation function should evolve via stable clustering (e.g., \citealt{Pea96}).  In the linear regime, clustering evolution can be parameterized by substituting $D^{(2+\gamma)}_z$ for $(1+z)^{-(3+\epsilon)}$ in Equation~\ref{eqn:realcorr}---$D_z$ is often called the linear growth factor.  For an $\Omega_m=\Omega_{total} = 1$ cosmology, $D_z = (1+z)^{-1}$.  For flat, $\Lambda$ cosmologies, $D_z$,  is suppressed as


\begin{equation}
D_z = \frac{g_z}{g_0}\frac{1}{(1+z)}
\end{equation}

\noindent where $g_0$ normalizes to the fiducial case, and $g$ may be approximated as

\begin{equation}
g_z \approx \frac{5}{2}\Omega_{mz}\left[\Omega_{mz}^{4/7}-\Omega_{\Lambda z}+\left(1+\frac{\Omega_{mz}}{2}\right)\left(1+\frac{\Omega_{\Lambda z}}{70}\right)\right]^{-1}
\label{eqn:gapprox}
\end{equation}


\noindent \citep{Car92}. The cosmological parameters (in a flat cosmology) evolve as


\begin{equation}
\Omega_{mz} = \left[\frac{H_0}{H_z}\right]^2\Omega_m(1+z)^3~~~,~~~\Omega_{\Lambda z} =\left[\frac{H_0}{H_z}\right]^2\Omega_\Lambda
\end{equation}

\noindent Using Equation~\ref{eqn:gapprox} is 30-40 times faster than integrating 
the cosmological parameters and produces a $g_z/g_0$ ratio consistent with Equation 28 of \citet{Car92} to 0.2\% or better for all redshifts (in our chosen cosmology). In the linear formalism, the de-projection can be derived by substituting $D_z^2$ for $(1+z)^{\gamma-(3+\epsilon)}$ in Equation~\ref{eqn:deproj}.

\begin{figure}
\plotone{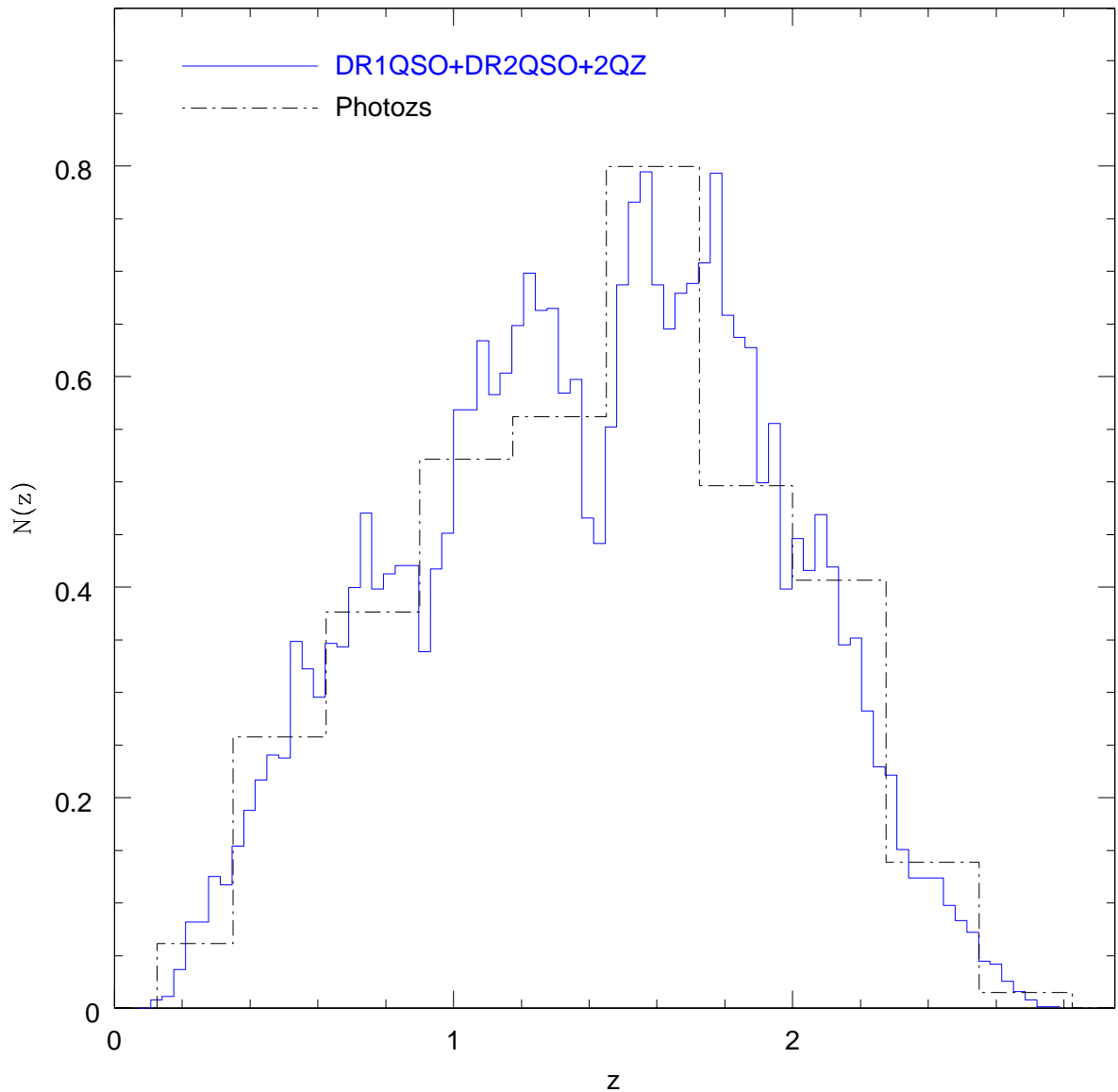}
\caption{The (normalized) redshift distributions for KDE objects (with $A_g < 0.18$).  The solid line combines the redshifts of spectroscopic matches to the KDE catalog from the 2QZ, DR1QSO and DR2.  Any objects that appear in multiple catalogs are assigned a redshift in the order DR1QSO-2QZ-DR2. The dot-dash line is the histogram returned assuming that the photometric redshifts in the KDE catalog are exact (binning is coarser to reflect imprecision in these estimates).\label{fig:zcomp}}
\end{figure}

Given an estimate of ${\rm d}N/{\rm d}z$, we can determine $r_0$, the local scale-length of the correlation function. KDE quasars are trained on DR1QSO colors, which are flux-limited in $i$, brighter than the $g < 21$ KDE limit.  However, the KDE redshift selection to $g < 21$ should resemble that of UVX-selected quasars as KDE objects are weighted against the stellar locus to $g < 21$ and then undergo a UVX cut. Ideally, we would obtain spectra for a small, random sample of KDE objects to establish selection but in the absence of this information our best sample consists of all known UVX quasars in the field. We therefore determine ${\rm d}N/{\rm d}z$ from all ($A_g < 0.18$) KDE objects with spectroscopic matches (in DR1QSO, DR2 or the 2QZ).  We have recreated our analysis using only the redshifts of DR1QSO matches and find our results are affected $\sim 1\%$.  Using only the redshifts of matches to DR2 quasars or to the 2QZ affects our results $\sim 3\%$. Such small changes are well within our random error.  In Fig~\ref{fig:zcomp} we plot the spectroscopic redshift distribution we use, in comparison to that obtained assuming the photometric redshifts of the KDE sample are exact. The two histograms are broadly consistent, which is entirely to be expected as, in an ensemble sense, the colors of the spectroscopic and photometric quasar samples are broadly consistent.

\begin{figure}
\plotone{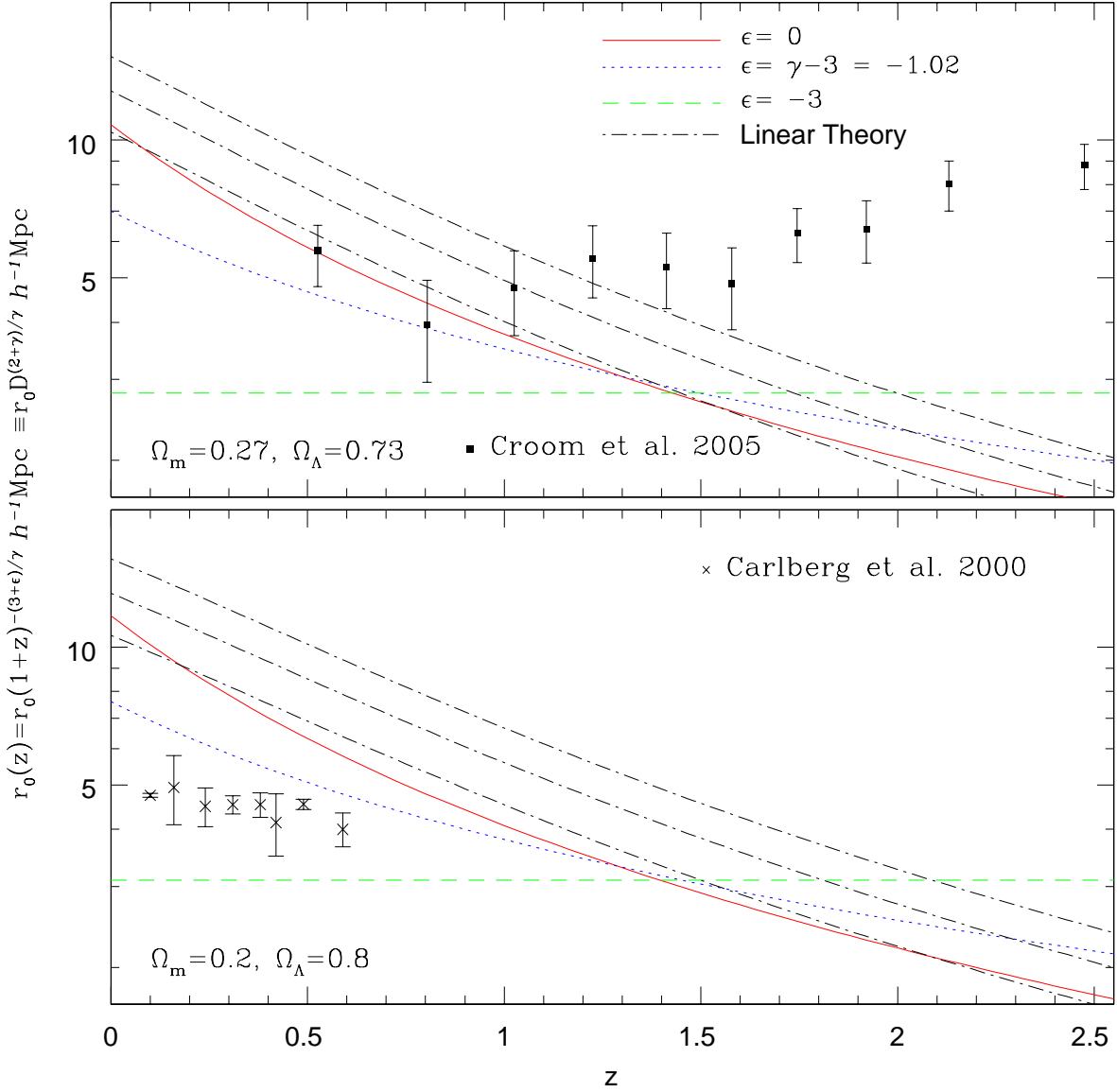}
\caption{Evolution of the real-space correlation scale length.  The models discussed in section~\ref{sec:Limb} are normalized to the best-fitting amplitude and slope for the KDE autocorrelation (the solid line in Figure~\ref{fig:extended}).  These models are compared to spectroscopic measurements of the quasar autocorrelation from \citet{Cro05} in the upper panel and the autocorrelation of bright galaxies from \citet{Car00} in the lower panel.  In both cases the cosmology is chosen to match the spectroscopic data.  The linear theory models are plotted with 1$\sigma$ error bars. Note that the $r_0(z)$ scale is logarithmic.
 \label{fig:evolution}}
\end{figure}

In Figure~\ref{fig:evolution}, we display the results of  de-projecting the solid line in Figure~\ref{fig:extended} to obtain $r_0$. In the upper panel of Figure~\ref{fig:evolution}, we plot $r_0(z)$, for a range of models in comparison to scale-lengths obtained for spectroscopically-confirmed quasars by \citet{Cro05}. In the lower panel of Figure~\ref{fig:evolution}, we compare to CNOC2 galaxies \citep{Car00}, as locally there is evidence that galaxies cluster like AGN \citep{Wak04}.  CNOC2 galaxies are luminous, so might be more clustered than average galaxies. Nevertheless, clustering of KDE objects and CNOC2 galaxies are most consistent for evolutionary models that predict little or no evolution in clustering.  Although our clustering measurements, being normalized at $z\sim1.4$, provide the strongest constraint when combined with local measurements, the \citet{Cro05} data in the upper panel of Figure~\ref{fig:evolution} further demonstrate that linear theory is increasingly inconsistent with quasar clustering at higher redshift. Linear theory predicts that dark matter, having had less time to collect under gravity, would be more clustered locally than at earlier times.  We find, in contrast, that QSOs are better represented by a model where their clustering is nearly constant with redshift from $z\sim1.4$ to $z\sim0$.  If linear theory is to be correct, this confirms a picture where QSOs were more clustered than underlying matter at high redshift, as has been argued for $\Lambda$CDM cosmogonies \citep{Efs88}.  Further, under the assumption of linear theory, the UVX quasar phase must be {\it short}, as is obvious from the dot-dash lines in the upper panel of Figure~\ref{fig:evolution}---if we could see the same quasars locally as at $z\sim1.4$ they would now be significantly more clustered than galaxies, having a scale-length of $12.8\pm2.4\Mpch$.

This analysis was carried out using simple assumptions; Limber's Equation, the angular distribution of KDE objects and the redshift distribution of spectroscopically-confirmed QSOs; however, it might be criticized for several reasons, most notably, QSO clustering, as we measure it, is probing a huge volume and being averaged across many different scales. \citet{Jen98} have found that dark matter in \LCDM simulations displays a local scale length of $r_0(z=0)\sim5\Mpch$, and has an approximate power-law form over scales of $1-20\Mpch$. If we restrict our analysis to these scales, then evolution of QSO clustering according to linear theory is still rejected at the $99.9$\% level, for a local scale length of $r_0=5\Mpch$.  Regarding this simple analysis as strong evidence that QSOs at high redshift were more clustered than the dark matter they traced, we will proceed by using photometric redshift information to quantify this evolution in terms of bias.

\subsection{Angular QSO clustering in Photometric Redshift Bins}
\label{sec:gausswide}

\citet{Kai84} first discussed biasing schemes, when conjecturing that rich clusters form where the clustering amplitude of dark matter exceeds some threshold. \cite{Bar86} extended the concept of bias to galaxies, or any object that formed in the rare peaks of a Gaussian random field.  Though bias might be a complex function of formation processes, it is often represented by a simple linear factor, $b$, that should acceptably parameterize QSO clustering relative to underlying dark matter (i.e., $\xi_{QQ} = b^2\xi$).

In the previous section, we modeled QSO clustering evolution via Equation~\ref{eqn:realcorr}, or it's linear theory equivalent, by assuming $\xi_{QQ} \equiv \xi$.  An alternative representation is

\begin{equation}
\frac{\xi_{QQ}(r,z)}{b_Q^2(r,z)} = \xi(r,z) = \left(\frac{r}{r_0(z)}\right)^{-\gamma}=\left(\frac{r}{r_0}\right)^{-\gamma}D_z^{(2+\gamma)}
\label{eqn:biascorr}
\end{equation}

\noindent allowing underlying dark matter to evolve according to linear theory even if $r_0(z)$ does not, as QSO clustering traces a bias parameter that may evolve with redshift. 

Equation~\ref{eqn:biascorr} can be de-projected as for Equation~\ref{eqn:realcorr}, but the results are now interpreted differently.  If we assume $r_0(z=0)$ = $5\Mpch$ for the matter correlation function, as is appropriate for local dark matter in $\Lambda$CDM simulations \citep{Jen98}, then

\begin{equation}
b_Q^z = \left(\frac{r_0^z}{5\Mpch}\right)^{\gamma/2}
\end{equation}

\noindent where the $z$ superscript indicates that these local values are implied by de-projecting from a particular redshift. Any scale-dependence in our measure of $b$ and $r_0$ can be ignored provided we average over scales large enough for linear theory to hold.  Applying this model to the linear theory scale length deduced in section~\ref{sec:Limb}, we find $b_Q^{z\sim1.4}=2.51\pm0.46$ for $\omega(\theta>2~{\rm arcmin})$. Of course this result is averaged over many different redshifts.

To better quantify the evolution of QSO bias, we redo the analysis of section~\ref{sec:Limb} in photometric redshift shells (see, e.g., \citealt{Bru00}).  Although diluting the statistics of the KDE data, this approach is attractive as narrowing the redshift range of any analysis allows more consistent scales to be de-projected.  Though the dispersion in the photozs, as compared to objects with a spectroscopic redshift, is typically $\Delta \sim 0.1-0.2$, causing scatter between bins, this dispersion should merely introduce noise, provided the redshift shells are large enough, i.e., larger than the typical $\Delta \zp$. Given that bin sizes are restricted by $\Delta \zp$, there is motivation to improve QSO photoz estimation.

An aspect of the photometric redshifts that might dilute clustering, by scattering quasars into entirely the wrong bin, is that some photoz estimates have likely solutions at several discrepant redshifts (these are sometimes called {\em catastrophic} estimates).  \cite{Wei04} quantify the probability of their photozs, so we could reduce our sample to those KDE objects that certainly do not have catastrophic redshift estimates.  However, after the previous cuts we made to remove stars, only 7\% of our sample have worse than a 50\% chance of being in their estimated photometric redshift interval, and most of these objects have several secondary solutions, rather than a single clear alternative.  We will thus discard no objects from our sample on the basis of photozs.  To test this, we have repeated our analyses with a randomly chosen 7\% of our sample assigned to a different photometric redshift bin, and find that results fluctuate negligibly on all scales, and well within the errors.  The likely reason why catastrophic estimates do not adversely impact our analysis is our large bin sizes ($\Delta \zp$), and there is thus motivation to reduce the number of catastrophic estimates while improving QSO photoz estimation.

\begin{figure}
\plotone{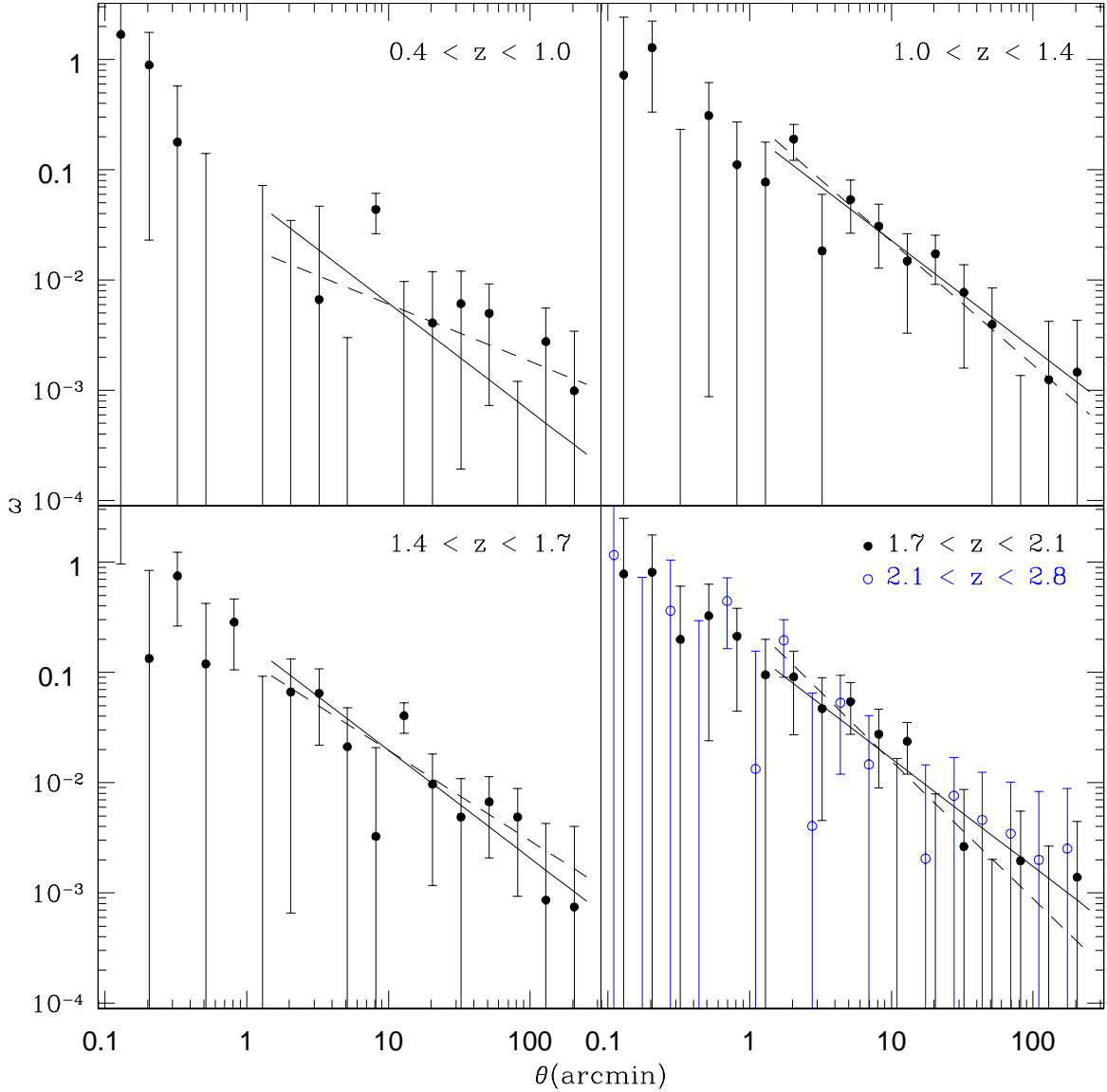}
\caption{The KDE autocorrelation in photometric redshift bins. In each bin, the solid circles are our standard $A_g<0.18$ KDE sample ($\sim$16,600 quasars per bin), the solid line is the best-fitting power-law with $\gamma=1.98$ (as determined in Figure~\ref{fig:extended}) and the dashed line is the best-fitting power-law with $\gamma$ allowed to float as a free parameter. In the lower-right panel, we also plot a high redshift bin, which is equivalent to everything with $\zp \geq 2.1$ from the KDE catalog---the open circles that represent these data are offset slightly.  The model fits to the open circles are not plotted, as they are very similar to those for the $1.7 < z_{phot} < 2.1$ bin (see Table~\ref{table:1}). All errors in this plot are jackknifed.
 \label{fig:zbins}}
\end{figure}

We split our ($A_g < 0.18$) KDE sample into 4 photometric redshift bins, containing $\sim16,500$ objects each, and calculate the angular autocorrelation in each bin (solid circles in Figure~\ref{fig:zbins}). We limit our main analysis to the range $0.4 <  \zp < 2.1$, for several reasons. The data are UVX-selected, meaning the range $0.4 <  z < 2.3$ is approximately most sensitive to the QSO SED. \citet{Wei04} suggest that their photoz estimation is best at $0.8 < \zp < 2.2$ but we include lower redshift data in a bin from $0.4 <  \zp < 1.0$ to increase the number of objects for our analysis (and to help compare quasar clustering to galaxy clustering). Dispersion in redshifts for $0.4 <  \zp < 0.8$ is larger than at other redshifts but not by enough to preferentially scatter objects to $\zp > 1.0$. To avoid probing too many scales in a single redshift bin we use a scheme that conveniently samples both similar scales and similar numbers of objects.  As there remain significant numbers of $A_g < 0.18$ KDE objects (10882) at $\zp > 2.1$, we measure their autocorrelation to provide additional constraints at high redshift (open circles in Figure~\ref{fig:zbins}).


We consider fits to our data in Figure~\ref{fig:zbins} out to 14, 22, 35, 56 and $89\Mpch$. In general, we do not fit on scales $<0.75\Mpch$ (at the median bin redshift), which should be in the fully non-linear regime.  Given that \cite{Cro05} find no evolution in the autocorrelation slope, we fit a single $\gamma$ for all redshifts.  We use the slope at $\theta>2$~arcmin displayed as the solid line in Figure~\ref{fig:extended}.  We also allow $\gamma$ to float as a free parameter, to demonstrate that there is sufficient degeneracy between measurements of slope and amplitude that fixing the slope does not unduly fix our results. When calculating $b_Q$, we assume our autocorrelations trace the underlying matter correlation slope (but not the amplitude) and that the matter correlation has $r_0(z=0)= 5\Mpch$. Both assumptions are reasonable as the bias necessary to make \LCDM match the galaxy autocorrelation is nearly linear (and certainly $<1.1$ on scales $<20\Mpch$; \citealt{Jen98}), and $r_0\sim 5\Mpch$ for galaxies (e.g., \citealt{Bau96}). 

\begin{figure}
\plotone{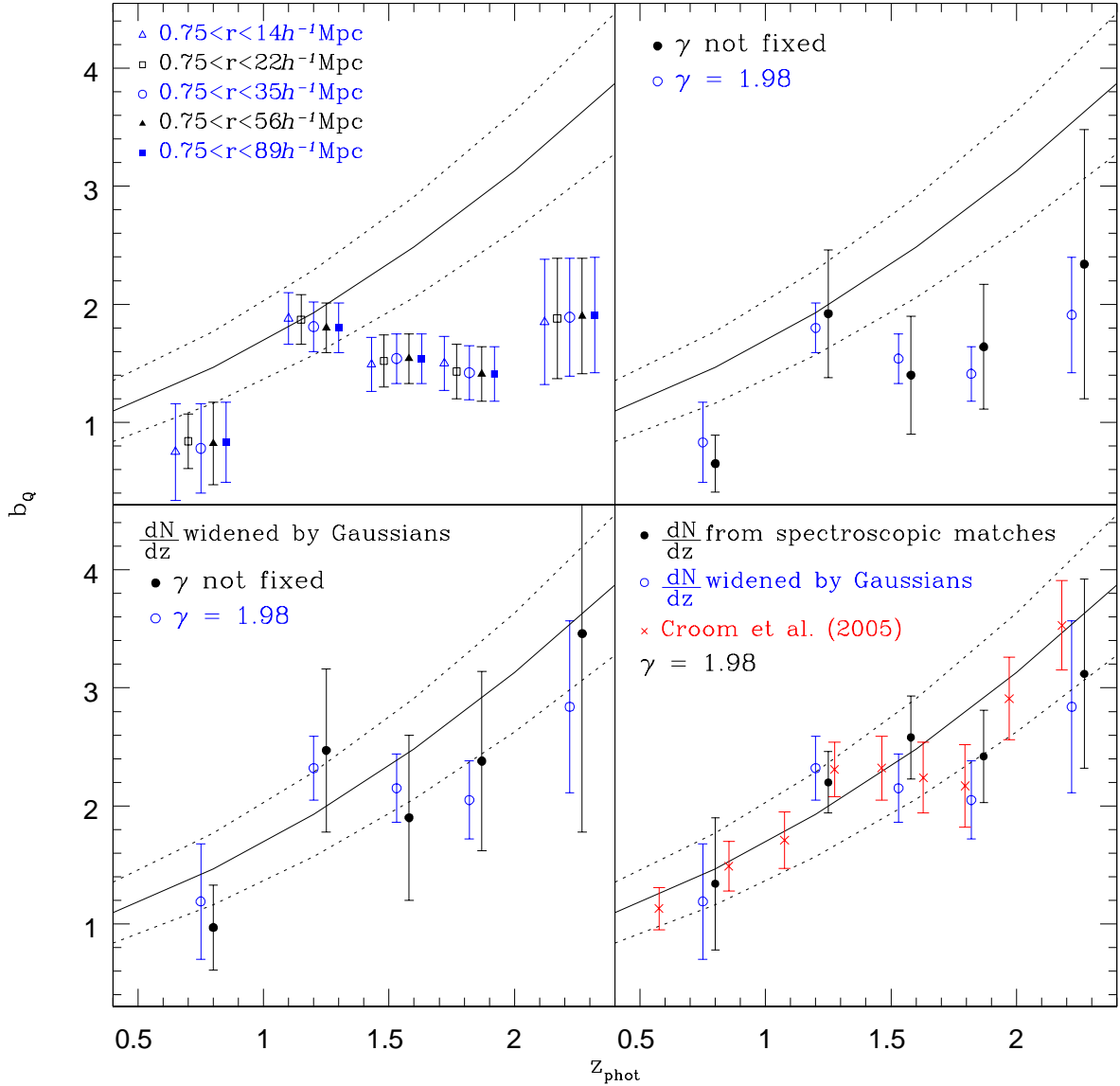}
\caption{Derived QSO bias evolution as a function of photometric redshift. The upper-left panel shows that estimates are consistent irrespective of the scale over which the correlation function is fit.  The upper-right panel demonstrates that the results are consistent whether the slope is fixed at $\gamma=1.98$ (the best-fit slope at $r > 0.75\Mpch$ for the full sample) or allowed to be a free parameter. The lower left panel mimics the upper-right but demonstrates the effect of widening the redshift distributions by Gaussians (see Equation~\ref{eqn:gauwide}). The lower-right panel compares the effect of widening the distribution by Gaussians to an alternate approach, determining d$N$/d$z_{phot}$ from spectroscopic matches in each $z_{phot}$ bin. The crosses are data from \citet{Cro05} and the solid lines plot $b_Q(z) = 0.53+0.289(1+z)^2$ (the empirical fit they derived from these data) with 1$\sigma$ error ranges (the dashed lines).
\label{fig:bias}}
\end{figure}

When de-projecting the correlation function we correct for imprecise photozs via the method of \citet{Bru00}.  After splitting our sample into photometric redshift bins, we widen the derived d$N/$d$\zp$ by two one-tailed Gaussians

\begin{equation}
\frac{{\rm d}N}{{\rm d}z} = [z_1,z_2]e^{-\left\{\left(z-[z_1,z_2]\right)/\sigma\right\}^2} ~;~[0 \leq z < z_1,z > z_2]
\label{eqn:gauwide}
\end{equation}

\noindent affixed at either end of a bin of $z_{phot1} \leq \zp \leq z_{phot2}$. Here, $\sigma$ is the dispersion between photometric and spectroscopic redshifts, which we derive from KDE objects with spectroscopic matches (in DR1QSO, DR2 or the 2QZ), after a $2\sigma$ clip to remove catastrophic photozs.  As a check on this method, we repeat our analysis assuming d$N/$d$z$ in each bin of $\zp$ is given simply by all spectroscopic matches to KDE quasars (see Table~\ref{table:1} and Figure~\ref{fig:bias}).

In Table~\ref{table:1}, we catalog the best-fitting models in photometric redshift bins.  Also shown are the de-projected scale lengths for the real-space correlation function (assuming the matter autocorrelation evolves according to linear theory), and the derived quasar bias (assuming the matter correlation has a local scale-length of $r_0(z=0)=5\Mpch$).  When the slope is fixed, and an appropriate d$N/$d$z$ is used, a model where $b_Q$ is both constant with redshift and linear (i.e. $b_Q=1$), is ruled out at high significance ($>99.9$\%, consistent with section~\ref{sec:Limb}).  Note that any small error in the $5\Mpch$ scale-length assumed for local dark matter would not affect our conclusions, as consistently changing $r_0$ introduces a systematic offset in $b_Q$ rather than noise.


In Figure~\ref{fig:bias}, we compare values of $b_Q$ from Table~\ref{table:1} to data from \citet{Cro05}.  We find consistent results irrespective of the scale we fit (upper left panel of Figure~\ref{fig:bias}). As the slope and amplitude of a power-law fit are degenerate, fixing $\gamma$ is not inconsistent with allowing $\gamma$ to vary as a free parameter. Allowing $\gamma$ to vary merely increases the error in our measurements of $b_Q$ and our results for $\gamma=1.98$ tend to the lower end of this increased error range, particularly at high redshift (top-right panel). Our data initially seem systematically lower than \citet{Cro05}; however, when the redshift distributions are widened by Gaussians, to reflect their photometric nature, $b_Q$ increases (lower-left panel). This technique of widening by Gaussians is consistent with determining d$N/$d$z$ from spectroscopic matches in each bin of photometric redshift (lower-right panel) and in either case our results, derived using photometrically-classified QSOs, are consistent with the results of \citet{Cro05}, which were derived by directly measuring the real-space clustering of (a smaller sample of) spectroscopically-confirmed QSOs.  Power-law fits to the data are reasonably acceptable at every redshift.  For the bins with $\bzp$=[0.75, 1.20, 1.53, 1.82, 2.22], respectively, $P(<\chi^2)$ = [0.34, 0.76, 0.65, 0.83, 0.96] for the fits out to $89\Mpch$.  Allowing $\gamma$ to vary as an extra parameter does not improve these fits, yielding $P(<\chi^2)$ = [0.34, 0.72, 0.62, 0.86, 0.94] 


The mean quasar luminosity increases with redshift for a flux-limited sample. However, as discussed by \citet{Cro05}, UVX quasars track characteristic quasar luminosities up to $z\sim2-2.5$.  For example, for the redshift shells used; $\bzp=[0.75$, $1.20$, $1.53$, $1.82$, $2.22]$; the mean absolute magnitude (calculated as in section~\ref{sec:lumclus}) is; $\bar{M_g} =[-22.57$, $-23.63$, $-24.17$, $-24.49$, $-25.02].$  At these redshifts, a ``characteristic luminosity'' for quasars (e.g., \citealt{Cro04,Ric05}) can be estimated as $M_g^* = [-23.81$, $-24.74$, $-25.23$, $-25.53$, $-25.75]$.  Thus the quasars we have considered are consistently around a magnitude fainter than $M_g^*$.  Ideally, in a large enough sample, we would separate the effects of luminosity from those of evolution by studying the quasar autocorrelation as a bivariate function of both luminosity and redshift.  In the next section, we attempt this, testing if it is feasible to measure the evolution of QSO clustering as a function of intrinsic luminosity.

\subsection{Quasar Clustering as a Function of Luminosity and Photometric Redshift}
\label{sec:lumclus}

To determine absolute magnitudes ($M_g$) for KDE objects, we assume each photoz is, on average, a good estimate of redshift and use it to calculate $M_g$ from $g$-band magnitudes.  We use a K-correction of $K(0.4 < z < 2.2)=-0.42-2.5(1+\alpha)\log(1+z)$, with (spectral index) $\alpha=-0.45$ (i.e., $f_\nu \propto \nu^\alpha$) from \citet{Wis00}, whom adopt a break at $z\sim0.4$ to ensure $K$ is zero locally, and suggest that this better approximates $K(z)$ at high redshift. Ideally, we would study quasar bias in equal bins of $M_g$, however, the large volume probed by quasars means that $M_g$ spans $\sim$8 magnitudes, and high and low redshift bins do not overlap.  It is thus difficult to fairly compare redshift bins without considering samples so small that noise dominates.  Instead, we split the KDE sample into three photoz bins that contain equal numbers, then subdivide these bins into three in $M_g$ that contain equal numbers.  We then measure the autocorrelation of each of these nine subsamples.

\begin{figure}
\plotone{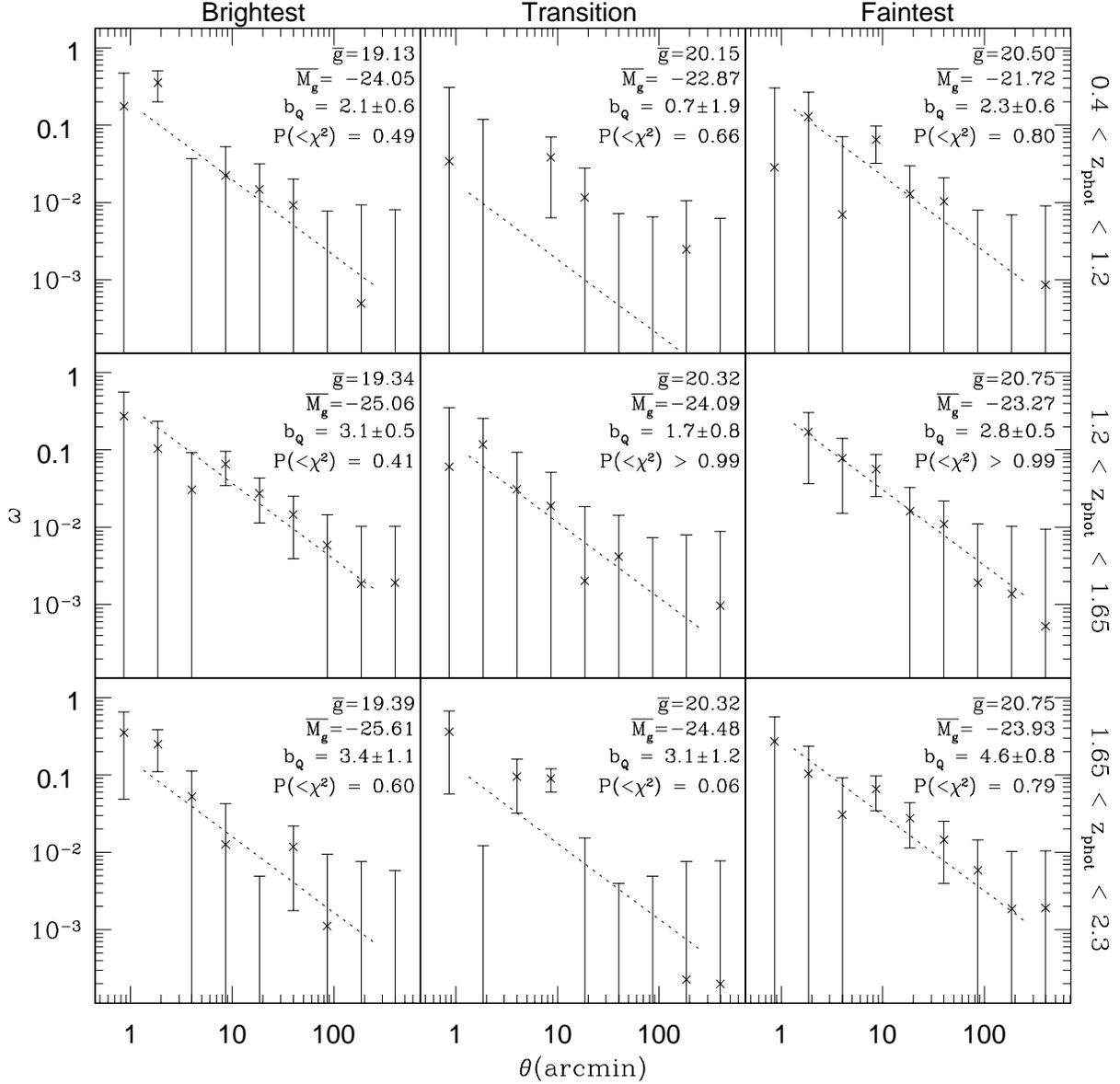}
\caption{Bivariate KDE QSO autocorrelation by redshift ($z_{phot}$) and absolute magnitude ($M_g$).  Our standard $A_g<0.18$ KDE sample is split into 3 redshift bins containing equal numbers and then further split into three $M_g$ bins of equal numbers (resulting in $\sim$8100 quasars per bin). The labels show the mean $g$ apparent magnitude, mean $M_g$, implied QSO bias ($b_Q$) and $\chi^2$ likelihood for each autocorrelation fit, while the dotted lines show the best-fit power law.  Low $\chi^2$ likelihoods could be improved by rebinning outliers without changing the fit. Errors in this plot are jackknifed.
\label{fig:absmag}}
\end{figure}

Figure~\ref{fig:absmag} shows $b_Q$ as a bivariate function of $M_g$ and redshift, derived as in section~\ref{sec:gausswide}, assuming $\gamma=1.98$ and using Equation~\ref{eqn:gauwide} for ${\rm d}N/{\rm d}z$. The implied values of $b_Q$ are noisy but are marginally consistent at every redshift {\it irrespective of absolute magnitude}. We therefore certainly cannot rule out the hypothesis that QSO clustering is independent of QSO luminosity.  Also, It appears that evolution of the quasar population with redshift has a stronger affect on QSO clustering than changes in quasars' luminosity; and, as would be expected from section~\ref{sec:gausswide}, we again find that QSOs are increasingly more biased with redshift, although this is rendered marginal by the lower numbers of objects in each bin.

\section{Discussion}

Implications of increasing QSO bias with redshift, and, more specifically, of the data plotted in the lower-right panel of Figure~\ref{fig:bias}, have been discussed by \citet{Cro05}, who use the ellipsoidal collapse model of \citet{She01} to estimate the mass of the dark matter halo (DMH) in which quasars of given bias reside.  \citet{Cro05} find that UVX QSOs reside in halos of similar mass at every redshift (see also; \citealt{Por04,Gra04}), and quote this mass as $M_{DMH} = (3.0 \pm 1.6)\times10^{12}h^{-1}M_\sun$.  This derived $M_{DMH}$ is locally consistent with the mass of an unbiased halo, $M^{*}$ (i.e. QSOs are locally biased similarly to $L^{*}$ galaxies).  A constant $M_{DMH}$ for quasars can be understood if $M^{*}$ and quasar bias evolve in step, so that QSOs are more biased at high redshift when $M^{*}$ is less massive.

Interestingly, the fact that $M_{DMH}$ for UVX QSOs does not evolve but that simulated dark matter halos merge and grow, suggests that objects observed in a QSO phase at high redshift must inhabit more massive dark matter halos by the present, and have turned off (or, more accurately, are no longer observed in UVX surveys, so are no longer in a UVX QSO phase).  Thus the QSO phase cannot be long-lived.  We are left with a picture where objects pass through a QSO phase when they inhabit dark matter halos of a certain mass.  A likely scenario is that the UVX quasar phase is triggered by a merger between halos of a characteristic mass (or two galaxies embedded in a single halo of that characteristic mass), and then is limited by some process to a timescale shorter than that typical of additional mergers, which would create a more massive parent halo for the quasar.  Meanwhile, smaller halos merge to eventually form more massive objects that harbor the correct conditions to ignite a QSO.  As the UVX quasar phase is a limited one, we predominantly see UVX quasars at a time ``close'' to the merger that triggered them, and certainly, on average, before further mergers take place---hence we see UVX quasars in a single, average halo mass. Further, the black holes that fueled quasars visible at redshifts of 2 and above should now reside in massive halos, they should also, however, have no UVX accretion signature.

Any short-lifetime model of the QSO phase must include a mechanism that can damp (UVX) quasar accretion processes on a timescale shorter than the typical merger rate.  An obvious explanation is a natural limit to quasar fuel reserves, or a central engine that grows to a point where it produces sufficient radiation pressure to expel its fuel source (e.g., \citealt{Sil98,Fab99,Saz05,DiM05}).  There is evidence \citep{Hut92,Mih96,Bah97} that quasars result from a merger of two gas-rich galaxies of near equal mass. Recent physical models \citep{Hop05a,Hop05b} demonstrate how such galaxy mergers could lead to quasars with a brief peak in optical luminosity, and a short-lived UVX stage.  If the mergers that cause UVX quasar activity occur between particularly massive, particularly gas-rich galaxies, then further mergers, once fuel is depleted in the Universe, would not necessarily reignite the UVX quasar phase (e.g., \citealt{Hop05c}). This could explain why, on average, additional mergers fail to produce populations of UVX quasars residing in more massive halos at low redshift.  However, models of galaxy formation must still explain {\it why} those mergers that initially trigger UVX quasars always occur within a halo (or two merging halos) with a total characteristic mass $\sim 3\times10^{12}h^{-1}M_\sun$ (averaged over the population at a given redshift), and why mergers that occur in more (or less) massive halos do not produce quasars typically observed in UVX surveys.  In the self-consistent galaxy-merger approach of \citet{Hop05c} this characteristic mass is self-evident as the mass of those AGN that have a peak optical luminosity above the observable threshold.  However, the luminosity distribution of quasars is empirically set in this model, so the equivalent question is why galaxy mergers in the Universe have led to quasars with the observed distribution of peak luminosities.

We find that the luminosity of the QSO population at a given epoch bears no significant relationship to QSO bias, in agreement with results from spectroscopic surveys (see, e.g., \citealt{Cro05}).  As luminosity-independent QSO bias would suggest the mass of a quasar's parent halo is independent of the quasar's luminosity, the implications of such a relationship merit speculation.  Given the $M_{BH}-\sigma$ correlation \citep{Fer00,Geb00} it is unlikely that the masses of the black holes ($M_{BH}$) that drive QSOs bear no correlation with the masses of the halos they reside in---but there could be a range of accretion efficiencies across the QSO population, with more luminous QSOs at every redshift having more efficient accretion process.  Again, this schema naturally arises within the formalism of \citet{Hop05c}, and \citet{Lid05} have predicted the implications of that model, finding them to be consistent with the empirical fit from \citet{Cro05} that we plotted in each panel of Figure~\ref{fig:bias}. The models applied by \citet{Lid05} formally predict little or no luminosity dependence to quasar clustering, again broadly consistent with the work in this paper.  However, we note that our data are currently sparse and noisy, and it remains to be seen definitively, from larger QSO samples, whether there is any luminosity dependence to the clustering of QSOs. We are engaged in producing a larger sample of photometrically-classified QSOs that we hope will formally start to constrain any quasar clustering as a function of luminosity.



\section{Conclusions and Future Work}

We have used a large sample of photometrically-classified QSOs from DR1 to estimate the quasar autocorrelation as a function of luminosity and photometric redshift.  We have demonstrated similar results using our ``proof-of-concept'', photometrically-classified sample as were obtained with the largest statistically-defined, spectroscopically-confirmed QSO sample available contemporaneously with DR1. Using photometric redshifts, we have confirmed that quasar clustering shows little evolution, suggesting QSOs are increasingly more biased with redshift up to $\bar{z}\sim2.2$.  We have attempted to measure the bivariate clustering amplitude of QSOs, finding that in a given redshift range there is no measurable dependence of QSO bias on QSO luminosity, and that evolution of the quasar population with redshift seems to have a stronger effect on QSO bias than changes in quasars' luminosity---however the errors on this measurement are still large enough to allow
some overlap between redshift evolution and luminosity evolution, and a larger data sample (or improved techniques) will be necessary to make definitive conclusions. We speculate that QSO luminosity evolution is likely independent of mass, depending mainly on accretion efficiency, and have discussed the implications of this in light of the recent models of \citet{Hop05a,Hop05b,Hop05c}.

We have confirmed that, for $A_g < 0.18$, the KDE catalog of \citet{Ric04} is contaminated by stars at only the 5\% level, easily low enough that quasar clustering can be meaningfully studied.  We have only used a fraction ($\sim$20\%) of the eventual SDSS data, so the KDE technique should prove an impressive resource for quasar science.  We note that there is room for improvement in the efficiency of the KDE algorithm at fainter magnitudes. There are firm scientific reasons for improving the classification of faint QSOs, not least of which is testing the evolution of QSO clustering across a significant range of QSO luminosity.  Larger photometric samples and improved faint-end classification will increase the overall number of photometrically-classified QSOs, which will both improve the significance of multivariate estimation and facilitate quasar autocorrelation estimation with improved angular binning resolution.  Larger samples and better classification will not, alone, be sufficient to increase binning resolution in redshift space beyond that used in this paper, as the typical photoz dispersion is currently comparable to the bin size.  It is thus useful and necessary to also improve QSO photometric redshift estimation.  We are currently engaged in producing a larger catalog of photometrically-classified quasars using improved techniques with novel priors and believe that QSO clustering measurements will soon be repeated with $\sim$500,000 quasars that are even more efficiently classified.

\acknowledgments

ADM and RJB wish to acknowledge support from NASA through grants NAG5-12578 and NAG5-12580 as well as support through the NSF PACI Project. GTR was supported in part by NSF grant AST-0307409 and a Gordon and Betty Moore Fellowship in Data Intensive Science. DVB and DPS acknowledge NSF support through grant AST03-07582. RCN acknowledges the EU Marie Curie Excellence Chair for support during this work. The authors made extensive use of the storage and computing facilities at the National Center for Supercomputing Applications and thank the technical staff for their assistance in enabling this work. We thank the referee, Scott Croom, for several invaluable contributions.

Funding for the creation and distribution of the SDSS Archive has been provided by the Alfred P. Sloan Foundation, the Participating Institutions, the National Aeronautics and Space Administration, the National Science Foundation, the U.S. Department of Energy, the Japanese Monbukagakusho, and the Max Planck Society. The SDSS Web site is http://www.sdss.org/. 

The SDSS is managed by the Astrophysical Research Consortium (ARC) for the Participating Institutions. The Participating Institutions are The University of Chicago, Fermilab, the Institute for Advanced Study, the Japan Participation Group, The Johns Hopkins University, the Korean Scientist Group, Los Alamos National Laboratory, the Max-Planck-Institute for Astronomy (MPIA), the Max-Planck-Institute for Astrophysics (MPA), New Mexico State University, University of Pittsburgh, University of Portsmouth, Princeton University, the United States Naval Observatory, and the University of Washington.

\clearpage

\clearpage

\clearpage

\clearpage

\clearpage

\clearpage

\clearpage

\clearpage

\clearpage

\clearpage

\begin{deluxetable}{crrrrr}
\tabletypesize{\scriptsize}
\tablecaption{Estimates of the QSO bias, $b_Q$, from the amplitude of the de-projected correlation function in photometric redshift bins (with median redshift $\bzp$).\label{table:1}}
\tablewidth{0pt}
\tablehead{
 Method/Scale at $\bzp$ bin & \colhead{$\bzp = 0.75$} & \colhead{$\bzp = 1.20$} & \colhead{$\bzp = 1.53$} & \colhead{$\bzp = 1.82$} & \colhead{$\bzp = 2.22$} }
\startdata
\sidehead{Observed Amplitude, $A \left({\rm arcmin}^{1-\gamma}\right)$\tablenotemark{a}}
$0.75 \leq r < 14~h^{-1}~{\rm Mpc}^{-1}$ & 0.049 $\pm$ 0.053 & 0.238 $\pm$ 0.055 & 0.177 $\pm$ 0.055 & 0.180 $\pm$ 0.054 & 0.145 $\pm$ 0.082 \\
$0.75 \leq r < 22~h^{-1}~{\rm Mpc}^{-1}$ & 0.060 $\pm$ 0.051 & 0.235 $\pm$ 0.053 & 0.186 $\pm$ 0.053 & 0.162 $\pm$ 0.053 & 0.149 $\pm$ 0.080 \\
$0.75 \leq r < 35~h^{-1}~{\rm Mpc}^{-1}$ & 0.052 $\pm$ 0.050 & 0.220 $\pm$ 0.052 & 0.191 $\pm$ 0.052 & 0.161 $\pm$ 0.052 & 0.151 $\pm$ 0.079 \\
$0.75 \leq r < 56~h^{-1}~{\rm Mpc}^{-1}$ & 0.058 $\pm$ 0.049 & 0.218 $\pm$ 0.051 & 0.190 $\pm$ 0.052 & 0.157 $\pm$ 0.051 & 0.152 $\pm$ 0.079 \\
$0.75 \leq r < 89~h^{-1}~{\rm Mpc}^{-1}$ & 0.059 $\pm$ 0.049 & 0.218 $\pm$ 0.051 & 0.190 $\pm$ 0.052 & 0.158 $\pm$ 0.051 & 0.154 $\pm$ 0.079 \\
$\gamma$ not fixed\tablenotemark{b}\tablenotemark{e} & 0.020 $\pm$ 0.015 & 0.30 $\pm$ 0.17 & 0.13 $\pm$ 0.09 & 0.28 $\pm$ 0.18 & 0.39 $\pm$ 0.38 \\
\tableline
\sidehead{De-projected $r_0$}
$0.75 \leq r < 14~h^{-1}~{\rm Mpc}^{-1}$ & 3.8 $\pm$ 2.1 & 9.5 $\pm$ 1.1 & 7.5 $\pm$ 1.2 & 7.5 $\pm$ 1.2 & 9.3 $\pm$ 2.7 \\
$0.75 \leq r < 22~h^{-1}~{\rm Mpc}^{-1}$ & 4.2 $\pm$ 1.9 & 9.4 $\pm$ 1.1 & 7.6 $\pm$ 1.1 & 7.2 $\pm$ 1.2 & 9.4 $\pm$ 2.6 \\
$0.75 \leq r < 35~h^{-1}~{\rm Mpc}^{-1}$ & 3.9 $\pm$ 1.9 & 9.1 $\pm$ 1.1 & 7.7 $\pm$ 1.1 & 7.1 $\pm$ 1.2 & 9.5 $\pm$ 2.5 \\
$0.75 \leq r < 56~h^{-1}~{\rm Mpc}^{-1}$ & 4.1 $\pm$ 1.8 & 9.1 $\pm$ 1.1 & 7.7 $\pm$ 1.1 & 7.1 $\pm$ 1.2 & 9.6 $\pm$ 2.5 \\
$0.75 \leq r < 89~h^{-1}~{\rm Mpc}^{-1}$ & 4.2 $\pm$ 1.7 & 9.1 $\pm$ 1.1 & 7.7 $\pm$ 1.1 & 7.1 $\pm$ 1.2 & 9.6 $\pm$ 2.5 \\
d$N$/d$z$ from spectra\tablenotemark{b}\tablenotemark{c} & 6.7 $\pm$ 2.8 & 11.1 $\pm$ 1.3 & 13.0 $\pm$ 1.8 & 12.2 $\pm$ 2.0 & 15.8 $\pm$ 4.1 \\
d$N$/d$z$ widened by Gaussian\tablenotemark{b}\tablenotemark{d}& 6.0 $\pm$ 2.5 & 11.7 $\pm$ 1.4 & 10.8 $\pm$ 1.5 & 10.3 $\pm$ 1.7 & 14.4 $\pm$ 3.7 \\
$\gamma$ not fixed\tablenotemark{b}\tablenotemark{e} & 2.8 $\pm$ 1.4 & 9.3 $\pm$ 2.4 & 7.0 $\pm$ 3.0 & 7.7 $\pm$ 2.2 & 9.9 $\pm$ 3.8 \\
$\gamma$ not fixed, d$N$/d$z$ widened\tablenotemark{b}\tablenotemark{d}\tablenotemark{e} & 4.8 $\pm$ 2.4 & 11.74 $\pm$ 3.1 & 10.0 $\pm$ 4.0 & 10.8 $\pm$ 3.1 & 13.46 $\pm$ 5.23 \\
\tableline
\sidehead{Derived $b_Q$}
$0.75 \leq r < 14~h^{-1}~{\rm Mpc}^{-1}$ & 0.75 $\pm$ 0.41 & 1.88 $\pm$ 0.22 & 1.49 $\pm$ 0.23 & 1.50 $\pm$ 0.23 & 1.85 $\pm$ 0.53 \\
$0.75 \leq r < 2~h^{-1}~{\rm Mpc}^{-1}$ & 0.84 $\pm$ 0.35 & 1.87 $\pm$ 0.21 & 1.52 $\pm$ 0.22 & 1.43 $\pm$ 0.23 & 1.88 $\pm$ 0.51 \\
$0.75 \leq r < 35~h^{-1}~{\rm Mpc}^{-1}$ & 0.78 $\pm$ 0.38 & 1.81 $\pm$ 0.21 & 1.54 $\pm$ 0.21 & 1.42 $\pm$ 0.23 & 1.89 $\pm$ 0.50 \\
$0.75 \leq r < 56~h^{-1}~{\rm Mpc}^{-1}$ & 0.82 $\pm$ 0.35 & 1.80 $\pm$ 0.21 & 1.54 $\pm$ 0.21 & 1.41 $\pm$ 0.23 & 1.90 $\pm$ 0.49 \\
$0.75 \leq r < 89~h^{-1}~{\rm Mpc}^{-1}$ & 0.83 $\pm$ 0.34 & 1.80 $\pm$ 0.21 & 1.54 $\pm$ 0.21 & 1.41 $\pm$ 0.23 & 1.91 $\pm$ 0.49 \\
d$N$/d$z$ from spectra\tablenotemark{b}\tablenotemark{c} & 1.34 $\pm$ 0.56 & 2.20 $\pm$ 0.26 & 2.58 $\pm$ 0.35 & 2.42 $\pm$ 0.39 & 3.12 $\pm$ 0.80 \\
d$N$/d$z$ widened by Gaussian\tablenotemark{b}\tablenotemark{d}& 1.19 $\pm$ 0.49 & 2.32 $\pm$ 0.27 & 2.15 $\pm$ 0.29 & 2.05 $\pm$ 0.33 & 2.84 $\pm$ 0.73 \\
$\gamma$ not fixed\tablenotemark{b}\tablenotemark{e} & 0.65 $\pm$ 0.24 & 1.92 $\pm$ 0.54 & 1.40 $\pm$ 0.50 & 1.64 $\pm$ 0.53 & 2.34 $\pm$ 1.14 \\
$\gamma$ not fixed, d$N$/d$z$ widened\tablenotemark{b}\tablenotemark{d}\tablenotemark{e} & 0.97 $\pm$ 0.36 & 2.47 $\pm$ 0.69 & 1.90 $\pm$ 0.70 & 2.38 $\pm$ 0.76 & 3.46 $\pm$ 1.68 \\
\enddata

\tablenotetext{a}{Throughout this table, unless otherwise noted, the assumed slope is $\gamma=1.98$.}
\tablenotetext{b}{This fit uses the ``$0.75 \leq r < 89\Mpch$'' definition for the scale.}
\tablenotetext{c}{Assumes the redshift distribution from spectroscopic matches (to DR1QSO, DR2 or the 2QZ) in each $\zp$ bin.}
\tablenotetext{d}{Assumes the redshift distribution from photometric redshifts widened by Gaussians (see Equation~\ref{eqn:gauwide}). The measured dispersions between spectroscopic and photometric redshifts are $\sigma=0.221,0.114,0.154,0.118,0.287$ for each listed bin, respectively, after applying a $2\sigma$ clip to remove catastrophic photometric redshift estimates.}
\tablenotetext{e}{These fits assume no slope for $\gamma$, instead taking the best fit from the data.}

\end{deluxetable}

\clearpage

\end{document}